\documentclass[sigconf]{acmart} 
\AtBeginDocument{%
  }

\copyrightyear{2026}
\acmYear{2026}
\setcopyright{cc}
\setcctype{by}
\acmConference[CHI '26]{Proceedings of the 2026 CHI Conference on Human Factors in Computing Systems}{April 13--17, 2026}{Barcelona, Spain}
\acmBooktitle{Proceedings of the 2026 CHI Conference on Human Factors in Computing Systems (CHI '26), April 13--17, 2026, Barcelona, Spain}
\acmPrice{}
\acmDOI{10.1145/3772318.3790713}
\acmISBN{979-8-4007-2278-3/2026/04}

\usepackage{xcolor}
\usepackage{listings} 
\usepackage{soul} 
\usepackage{booktabs}

\newcommand{\systemName}[0]{{Poppins}}
\newcommand{\userStudyN}[0]{{17}}
\newcommand{\screenSurveyN}[0]{{14}}
\newcommand{\prolificStudyN}[0]{{205}}

\colorlet{punct}{red!60!black}
\definecolor{background}{HTML}{FFFFFF}
\definecolor{delim}{RGB}{20,105,176}
\colorlet{numb}{magenta!60!black}

\lstdefinelanguage{json}{
    basicstyle=\footnotesize\ttfamily,
    stepnumber=1,
    numbersep=8pt,
    showstringspaces=false,
    breaklines=true,
    backgroundcolor=\color{background},
    literate=
     *{0}{{{\color{numb}0}}}{1}
      {1}{{{\color{numb}1}}}{1}
      {2}{{{\color{numb}2}}}{1}
      {3}{{{\color{numb}3}}}{1}
      {4}{{{\color{numb}4}}}{1}
      {5}{{{\color{numb}5}}}{1}
      {6}{{{\color{numb}6}}}{1}
      {7}{{{\color{numb}7}}}{1}
      {8}{{{\color{numb}8}}}{1}
      {9}{{{\color{numb}9}}}{1}
      {:}{{{\color{punct}{:}}}}{1}
      {,}{{{\color{punct}{,}}}}{1}
      {\{}{{{\color{delim}{\{}}}}{1}
      {\}}{{{\color{delim}{\}}}}}{1}
      {[}{{{\color{delim}{[}}}}{1}
      {]}{{{\color{delim}{]}}}}{1},
}

\lstdefinelanguage{Markdown}{
    basicstyle=\ttfamily\footnotesize, 
}

\begin{document}

\title[Just-In-Time Objectives: A General Approach for Specialized AI Interactions]{Just-In-Time Objectives: A General Approach for\\ Specialized AI Interactions}


\author{Michelle S. Lam}
\orcid{0000-0002-3448-5961}
\affiliation{%
  \institution{Stanford University}
  \city{Stanford}
  \state{CA}
  \country{USA}
}
\email{mlam4@cs.stanford.edu}

\author{Omar Shaikh}
\orcid{0000-0003-1393-8041}
\affiliation{%
  \institution{Stanford University}
  \city{Stanford}
  \state{CA}
  \country{USA}
}
\email{oshaikh@stanford.edu}

\author{Hallie Xu}
\orcid{0009-0001-0158-5069}
\affiliation{%
  \institution{Stanford University}
  \city{Stanford}
  \state{CA}
  \country{USA}
}
\email{halliexu@stanford.edu}

\author{Alice Guo}
\orcid{0000-0002-3674-5886}
\affiliation{%
  \institution{Stanford University}
  \city{Stanford}
  \state{CA}
  \country{USA}
}
\email{azguo@stanford.edu}

\author{Diyi Yang}
\orcid{0000-0003-1220-3983}
\affiliation{%
  \institution{Stanford University}
  \city{Stanford}
  \state{CA}
  \country{USA}
}
\email{diyiy@cs.stanford.edu}

\author{Jeffrey Heer}
\orcid{0000-0002-6175-1655}
\affiliation{%
  \institution{University of Washington}
  \city{Seattle}
  \state{WA}
  \country{USA}
}
\email{jheer@uw.edu}

\author{James A. Landay}
\orcid{0000-0003-1520-8894}
\affiliation{%
  \institution{Stanford University}
  \city{Stanford}
  \state{CA}
  \country{USA}
}
\email{landay@stanford.edu}

\author{Michael S. Bernstein}
\orcid{0000-0001-8020-9434}
\affiliation{%
  \institution{Stanford University}
  \city{Stanford}
  \state{CA}
  \country{USA}
}
\email{msb@cs.stanford.edu}

\renewcommand{\shortauthors}{M.S. Lam, O. Shaikh, H. Xu, A. Guo, D. Yang, J. Heer, J.A. Landay, M.S. Bernstein}

\begin{abstract}
    Large language models promise a broad set of functions, but when not given a specific objective, they default to generic results.
We demonstrate that inferring the user's in-the-moment objective, then rapidly optimizing for that singular objective, enables LLMs to produce specialized tools, interfaces, and responses.
Our work introduces \textit{just-in-time objectives}, which model a user's goals to specialize LLM systems on the fly.
We contribute an architecture for automatically inducing such objectives by passively observing user behavior, then steering downstream AI systems through generation and evaluation against this objective.
Inducing just-in-time objectives (e.g., ``Clarify the abstract's research contribution'') enables automatic generation of tools, e.g., those that critique a draft based on relevant HCI methodologies, anticipate related researchers' reactions, or surface ambiguous terminology.
In a series of experiments on participants' own tasks, JIT objectives enable LLM outputs that achieve 66--86\% win rates over typical LLMs. In-person use sessions confirm that JIT objectives produce specialized tools that are unique to each participant and are rated as significantly higher quality than a standard LLM chat tool.

\end{abstract}


\begin{teaserfigure}
    \centering
    \includegraphics[width=\textwidth]{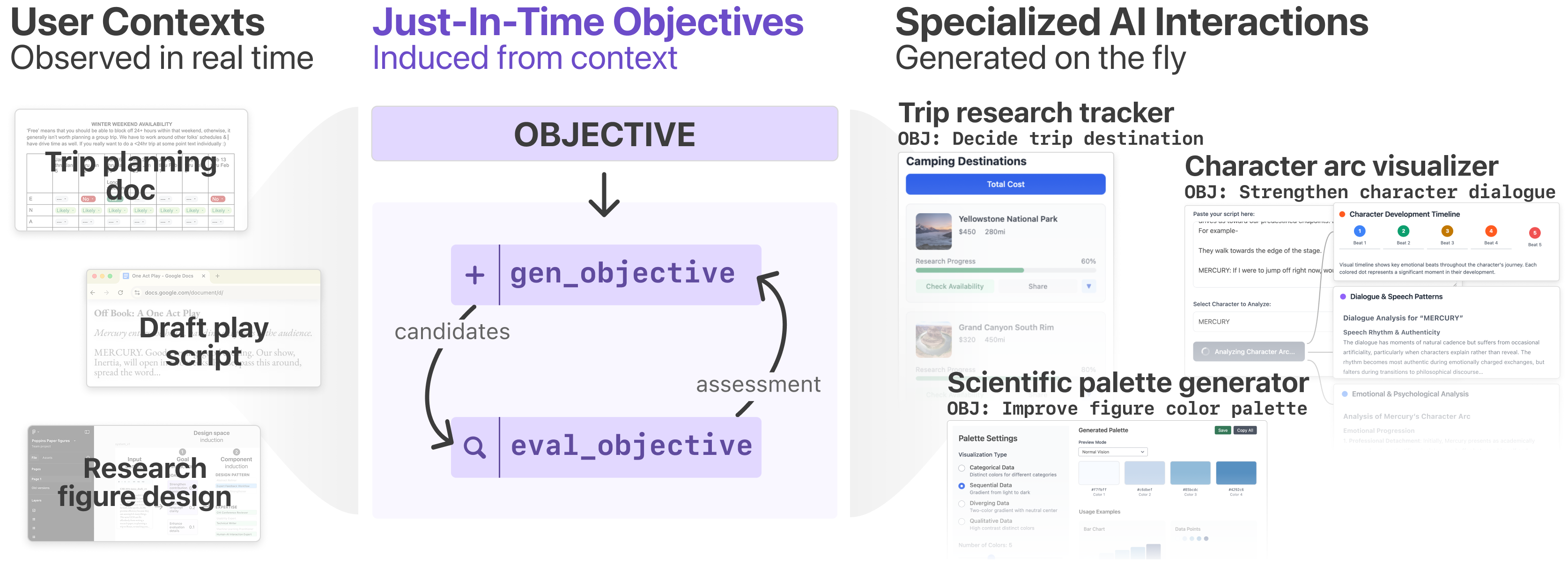}
    \caption{Just-in-time objectives are a method for producing AI objectives by inducing user goals on the fly from observations of the user and their task. We turn these objectives into first-class interactive objects that are visible, modifiable, and equipped to powerfully steer any number of downstream AI systems. By inducing possible objectives at interaction time, these objectives can be much more detailed and specialized to the particular task at hand compared to implicit, generic AI objectives.}
    \Description{A conceptual diagram showing the flow from user context observation to objective induction to specialized AI system outputs. The left side shows various windows of a user's workspace that have been passively observed. The center shows the just-in-time objectives that feed into generation and evaluation operations. The right side displays various AI-generated outputs including tools and interfaces that are more detailed and specialized compared to generic large language model outputs.}
    \label{fig:pull_figure}
\end{teaserfigure}


\maketitle

\section{INTRODUCTION}
In the course of writing this paper, we wondered what a large language model (LLM) might have to say about our introduction.
The model offered helpful syntax edits (``replace technical phrases like hill-climb''), suggestions to aid readability (``some sentences could be broken up''), and thoughts on re-ordering content (``consider moving implementation specifics to a later section''). These suggestions all constitute useful advice, but we found them rather milquetoast and unopinionated. Certainly a seasoned researcher would have more specific advice on our core argument. 

Why do we have such generic LLM output? One major factor is that training objectives for an LLM must be defined well in advance of the contexts where a model will be applied, so they must make assumptions about what users need.
Model post-training aims to make up for this lack of certainty by fine-tuning models against many simultaneous objectives, such as performance on reasoning tasks, value alignment, and preference data labeled by people with diverse opinions. However, like a design by committee, optimizing jointly over these objectives has the effect of converging the model to generic, safe outputs~\cite{sharma2024sycophancy, santurkar2023whose, chakrabarty2024artOrArtifice, shaikh2025ditto}. These generic objectives form a default that is rarely overridden even at interaction time: users often have difficulties envisioning model behavior~\cite{subramonyam2024gulfOfEnvisioning}, struggle with underspecified LLM prompts~\cite{zamfirescupereira2023whyJohnny, shaikh2024grounding, agrawala2023blackboxes}, and still end up converging to generic ideas~\cite{anderson2024homogenization,kumar2025humanCreativity}.

In this paper, we demonstrate that inferring the user's in-the-moment objective, then rapidly optimizing the LLM's behavior toward that singular objective, enables the LLM to generate outputs that are far more distinct and responsive: whether they be individual responses, extended drafts, or generated software tools.
Instead of relying on an anticipatory objective that must be generic and opaque by default, we find that we can pursue an on-demand objective that is specific and malleable by default.
User objectives over long stretches of time (e.g., a user's writing philosophy) are complex and challenging to anticipate, which causes standard preemptive LLM objectives to fail. But when objectives are created on-demand based on small windows of observed user behavior, they become more tractable and accurate (e.g., a user's aims for this phase of the Introduction).\footnote{To make an analogy to calculus, even if a user's objective is complex and curved over time, it can still be characterized as a simple straight line (a singular objective) over an infinitely small instant.}
We capture these instantaneous user goals as what we term \textbf{just-in-time objectives}: for example, ``clarify the calculus example in the introduction,'' ``restructure the argument logic,'' or ``respond to Reviewer 2's critique about sampling.''
By inducing possible objectives at interaction time, these objectives can be much more detailed and specialized to the particular task at hand compared to implicit, generic AI objectives.

We contribute a generalizable architecture for user-specific LLM specialization that automatically induces just-in-time (JIT) objectives from interaction traces, then applies the JIT objective to steer AI behavior by shaping LLM generation and evaluation. 
This architecture induces strong shifts in model behavior across task domains: we demonstrate that applying a JIT objective in a generate-then-rank architecture enables LLMs to rapidly construct tools that are specific and useful to the user in the moment. Recall that a standard LLM provided generic syntax and composition advice on an introduction like this one. Instead, applying an induced just-in-time objective of ``explain the system clearly in this CHI paper introduction'' enables a system to automatically produce drafts that rework this paragraph's logical flow to match that of similar introduction paragraphs from related CHI papers, or responses that simulate feedback from likely CHI reviewers, or tools that identify where the narrative veers away from describing the system. 

We instantiate our approach in a tool called \systemName{}: a browser extension and web application that observes user screens to infer just-in-time objectives, which allows the system to automatically produce detailed candidate tool designs and generate novel interactive tools to assist the user within minutes. 
The system captures the user's browser context, and uses that context to infer a just-in-time objective (e.g., ``enhance introduction draft by clarifying system architecture''). This objective steers \systemName{} to generate and evaluate candidate design specifications, then passes the best design specification to produce a functional software tool for the user. 

To understand whether our just-in-time objectives are both accurate and useful, we conduct a series of evaluations that assess the objectives in isolation and then incorporate them in increasingly complex LLM systems.
We first run experiments on participant-submitted browser traces collected over three days ($N=\screenSurveyN{}$) and find that across a variety of contexts in users' daily lives, our just-in-time objectives are accurate and useful (acc: $M=2.04$, useful: $M=2.18$ on a -3 to +3 Likert scale, with 75.0\% of ratings ``2: Accurate/Useful'' or higher). 
In addition, when applied to optimize LLM outputs that suggest feedback, areas of expertise, and tool designs, just-in-time objectives produce strongly-preferred outputs, achieving win-rates between 71\% and 86\% over those of a baseline LLM. 
These results are corroborated by a large-scale experiment ($N=\prolificStudyN{}$) across 410 participant-submitted input contexts, where just-in-time objectives are accurate, useful, and effective LLM optimizers (acc: $M=1.92$, useful: $M=2.06$, win-rates between 66\% and 71\% over baseline LLM). 
In hour-long study sessions applying \systemName{} to participants' own writing tasks ($N=\userStudyN{}$), we find that generated tools constructed using just-in-time objectives are relevant and useful without requiring manual prompting. \systemName{} outputs yield overall quality ratings significantly higher than a standard LLM chat tool ($p<.05$). \systemName{} produces a broad range of tailored tools including a ``Cultural Perspective
Highlighter'' for a scholarship personal essay, ``Neural Architecture Search (NAS) Explorer'' for a research project on microcontrollers, ``Technical Protocol Generator'' for a bioengineering research proposal, and ``Character Emotion Tracker'' for a science fiction short story draft.

Just-in-time objectives allow us to tackle the long tail of specific user needs while granting end users greater ability to shape their own tools. 
In summary, we contribute:
\begin{itemize}
    \item A generalizable architecture for \textit{just-in-time objectives}: a model of a user's in-the-moment goals that is automatically induced to achieve user-specific LLM specialization. Our method centers around an objective-based generation and evaluation loop to optimize AI output without supervision.
    \item The \systemName{}\footnote{Code available at: \url{https://github.com/michelle123lam/poppins}} system, which instantiates our vision of just-in-time objectives in the domain of generative user interfaces by translating observed user contexts to functional interactive tools. With \systemName{}, just-in-time objectives generate tools by inducing objective-aligned design specifications.
    \item Evaluations that validate the accuracy and utility of just-in-time objectives and outputs of the AI systems they optimize, such as expert feedback and generated software tools in \systemName{}.
\end{itemize}

\section{RELATED WORK}

Our work builds on evolving explorations of human-LLM interaction, as well as an emerging literature that names worrying individual- and population-level failure modes. Our approach works towards a goal of interactive systems that specialize on the fly, echoing classic literature on adaptive interfaces and recent work on user interface generation.

\subsection{Failure Modes of Human-LLM Interaction: Undirected Prompting and Generic Outputs}
Despite the excitement and hopefulness around large language models and chat-based interaction, recent work highlights critical cognitive barriers when interacting with LLMs with monolithic and opaque objectives. 
Ubiquitous prompt- and chat-based LLM user interfaces provide limited affordances to understand black-box LLMs, preventing users from forming reliable mental models to control system behavior~\cite{agrawala2023blackboxes}. Returning to the classic Gulfs of Execution and Evaluation~\cite{hutchins1985directManipulation}, related work locates new gulfs that may emerge for generative AI, such as a Process Gulf in understanding how AI executes tasks \cite{terry2023interactive} and a Gulf of Envisioning in setting the right goals in the first place when using AI~\cite{subramonyam2024gulfOfEnvisioning}. Natural language interaction is not a cure-all for human-AI interaction, but in fact leads to frustrations with underspecified prompts~\cite{shaikh2024grounding, zamfirescupereira2023whyJohnny}.
We are motivated by many of these cognitive barriers that arise from black-box LLMs. Our work argues that rather than relying on unstructured and noisy natural language requests, users may benefit from a more explicit construct of objectives that steer the model and transparently convey the model's current directive.

Another strong motivator of our work is the problem of generic LLMs.
Emerging research presents evidence of the homogenizing impacts of generic LLM outputs for both individual-level~\cite{padmakumar2024does, lee2025criticalThinking,kumar2025humanCreativity, chakrabarty2024artOrArtifice} and population-level creativity~\cite{anderson2024homogenization, wu2024generativemonoculture, doshi2024generative, wenger2025we, messeri2024artificial}.
From LLMs, we observe an uncanny valley of model behaviors that appear divergent, but are not fully random, and yet effortful to meaningfully steer~\cite{UganderEpstein2024Art}.
Worryingly, at a population level, LLMs appear to promote monocultures, steering users towards homogeneous, convergent thinking even when individual outputs appear creative~\cite{anderson2024homogenization, doshi2024generative}. 
One response is that we need pluralistic modeling approaches that better represent the diversity of the population~\cite{feng2024model, feng2024modular} and account for pluralistic values~\cite{sorensen2024value, sorenson2024pluralistic}.
Another response is that we need mechanisms for greater and more sustainable participation in the development and maintenance of LLMs~\cite{suresh2024participation, huang2024CCAI}. 
We share a concern about the risks of current LLM tools promoting monocultures, and our work explores whether even generic models can be effectively coaxed into specific, divergent variants with just-in-time objectives.
In contrast with approaches that specialize LLM behavior with domain- or task-specific context~\cite{wadhwa2025evalagents, lin2024wildbench}, our approach aims for user-specific specialization that can generalize across tasks and domains.
We posit that if we can grant users on-demand access to specific, divergent model behaviors, they may be able to break away from generic model outputs and actualize their unique creative voice.

\subsection{Adaptive UIs \& User Modeling: Building on Past Lessons}
Breaking from these LLM failure modes, our just-in-time objectives take an approach of observing the user and inferring their goals from their context, which is closely related to work on adaptive user interfaces and user modeling.
Adaptive interfaces have a long history in human-computer interaction, aiming to dynamically adjust functionality based on user context and inferred intent ~\cite{gajos2004supple, gajos2007supplePlusPlus, nichols2002PUC, nichols2006huddle, nichols2006uniform}. To set the right target for adaptation, many pioneering approaches have developed user models, for example to estimate user effort~\cite{gajos2004supple}, predict user motor capabilities~\cite{gajos2007supplePlusPlus}, or infer user goals or needs~\cite{horvitz1998lumiere,horvitz1999mixedInitiative}. We build on a similar pipeline to achieve adaptive system behavior by observing and making inferences about the user, but owing to the generative capabilities of LLMs, just-in-time objectives are not just models to aid selection over adaptive UI options, but the mechanism by which a system \textit{generates} adaptive UI options.

While in theory, adaptive interfaces seem unilaterally better than static interfaces, in practice, adaptive UIs face common pitfalls around unpredictability and loss of user control~\cite{findlater2009design}. With developments in LLMs, a number of recent works expand the vision of adaptive interfaces by supporting dynamic widgets and on-the-fly code execution in contexts such as data visualization and computational notebook tutorials~\cite{vaithilingam2024dynavis, cheng2024biscuit, dibia2023lida}.
While our work does not rid itself entirely of unpredictability due to the use of black-box LLMs, we argue that conjuring objective-aligned tools from a vast set of options (and making this objective visible) elides many of the traditional issues of adaptive UIs, which stemmed from a finite set of interactive functions moving around in unpredictable ways.

\subsection{Dynamic UI Generation: Direction-Setting With the Right Optimizer}

An emerging set of work explores dynamic UI generation, leveraging the generative flexibility of LLMs while providing stable UI affordances~\cite{litt2025malleable}. This work ranges from on-the-fly widgets for data visualization and computational notebooks~\cite{vaithilingam2024dynavis,cheng2024biscuit} to live behavior generation in games~\cite{jennings2024gromit}. While some work envisions generative user interfaces that evolve with flexible end-user customization~\cite{cao2025jelly, min2025malleable}, other work envisions them as widespread alternatives to status quo chat interfaces~\cite{chen2025generativeinterfaces}. 

We share an excitement about this research direction, while also noting that achieving a ``dynamic'' interface is not inherently valuable: interfaces must adapt along the axes that users find valuable. 
Recent work on UI generation demonstrates the importance of feedback to steer UI generation towards high-quality designs. Large-scale data such as eye-tracking data~\cite{jiang2024uEyes} and repositories of existing UIs~\cite{uicoder, wu2024uiClip} can generate reliable feedback, but such data is costly to gather.
We draw inspiration from self-improving algorithms for LLM pipelines~\cite{madaan2023selfrefine} that leverage lightweight metrics rather than massive datasets~\cite{khattab2024dspy, zhu2025autolibra}, but our work explores whether these metrics could be directly induced from passive observations of user behavior. 
Given such user-specific metrics, it is possible to build on the success of LLM generate-then-rank architectures that can scale up LLM performance at inference time with generators that produce a large number of samples and evaluators (or verifiers) that select the best output candidate~\cite{brown2024largeLanguageMonkeys, song2025mindTheGap, snell2024scaling, saadFalcon2025Weaver}.
Just-in-time objectives provide a means for users to express what axes they find valuable and guide UI generation, without requiring manual customization.

\section{JUST-IN-TIME OBJECTIVES}

With a critical role in shaping model behavior from conversational chatbots to code assistants to writing tools, large language model objectives are often fixed far in advance of the scenarios they will encounter. In-context prompting can overcome this issue, but users are rarely detailed enough in their request to cover all required decisions~\cite{shaikh2024grounding}. In this section, we describe an architecture that shifts the work of objective-making to the \textit{site of the user}, at the \textit{specific moment of need}.

Our work contributes a \textit{generalizable architecture} to achieve \textit{user-specific LLM specialization}:
\begin{itemize}
    \item Our primary goal is \textbf{user-specific LLM specialization}: LLM output that best aligns with a particular user’s needs. Domain- or task-specific needs can be gleaned from the web, expert-authored checklists, or model self-reflection as in prior work~\cite{madaan2023selfrefine, wadhwa2025evalagents, lin2024wildbench}. By contrast, user-specific needs depend on not just the domain (e.g., writing) or task (e.g., outlining), but factors such as the user’s specific task instance (e.g., the content, audience, and current progress of their outline), task environment (e.g., current application or tool, visible portion of outline), and personal preferences. All of these factors strongly shape a user’s current objective and can only be gathered from firsthand data about their specific interaction context.
    
    \item To achieve this goal, we contribute a \textbf{generalizable architecture} that functions across task domains. Ideas of user-specific specialization exist in individual domains (e.g., writing assistance, persona generation, web search \cite{benharrak2024writerDefined, dArcy2024MARG, choi2025proxona, baek2024knowledgeAugmented, salemi2024LaMP}), but are implemented with domain-specific assumptions and are not designed to transfer across domains. Our architecture can consistently and automatically apply user-specific specialization without domain-specific assumptions or manual steering from users or developers.
\end{itemize}

In the following sections, we walk through the technical architecture of just-in-time (JIT) objectives that can be implemented into any interactive system. We demonstrate how crisp objectives in-the-moment can shape AI applications, taking examples of common system components responsible for generation, evaluation, and iterative improvement.
We then illustrate how user-facing systems can be reimplemented with just-in-time objectives by introducing the \systemName{} system, which incorporates just-in-time objectives to drive customized UI generation.

\subsection{Architecture}

In an ideal world, a user-originated AI objective would come directly from the user requesting exactly what they want in natural language (like ``Clarify the motivation for the architecture''). This approach works well when users have clear goals and a willingness to communicate them to a system, but an emerging literature has documented users' frustrations in articulating their intentions to a model (``That's not what I mean by the architecture,'' ``This version is still not any clearer'')~\cite{zamfirescupereira2023whyJohnny, agrawala2023blackboxes}, or even knowing what they might articulate (``Is it worth asking the model to critique my draft, or should I just ask it for paper recommendations, or something else altogether?'')~\cite{subramonyam2024gulfOfEnvisioning}. 
While we believe users ought to have direct \textit{access} to author AI objectives, the right \textit{interface} for authoring AI objectives requires additional scaffolding beyond manual prompting.

Our architectural goal is that users should not need to start from scratch when communicating their objectives, and user-tailorable objectives should be enabled by default.
While there are many paths to achieve this, we demonstrate that by purely observing the user, even minimally via browser interaction logs or screenshots, our architecture can deploy calibrated propositions about a user (``User is working on a System section with comments from X, Y, and Z'')~\cite{shaikh2025gum}, which we translate into a working hypothesis of the user's objective at any moment in time~(``Iterate on System section by integrating feedback from collaborators'').

In this section, we work backwards by first demonstrating (1)~how such an objective can be minimally applied to common components of LLM systems to steer their behavior, and then discuss (2)~a process for creating just-in-time objectives by observing user behavior.
In particular, our architecture builds on the existing design pattern of generators and evaluators in AI systems. These are also known as generators and verifiers~\cite{saadFalcon2025Weaver, brown2024largeLanguageMonkeys, song2025mindTheGap, snell2024scaling} or generate-then-rank methods ~\cite{zhang2023autoInstruct, xu2020generateandrank}, with analogues to actor-critic~\cite{konda1999actorCritic} and generator-discriminator architectures~\cite{goodfellow2014GANs} for prior AI systems. We design our system so that induced objectives can slot into existing generators and evaluators in LLM systems.

\begin{figure*}[!tb]
    \centering
    \includegraphics[width=0.55\linewidth]{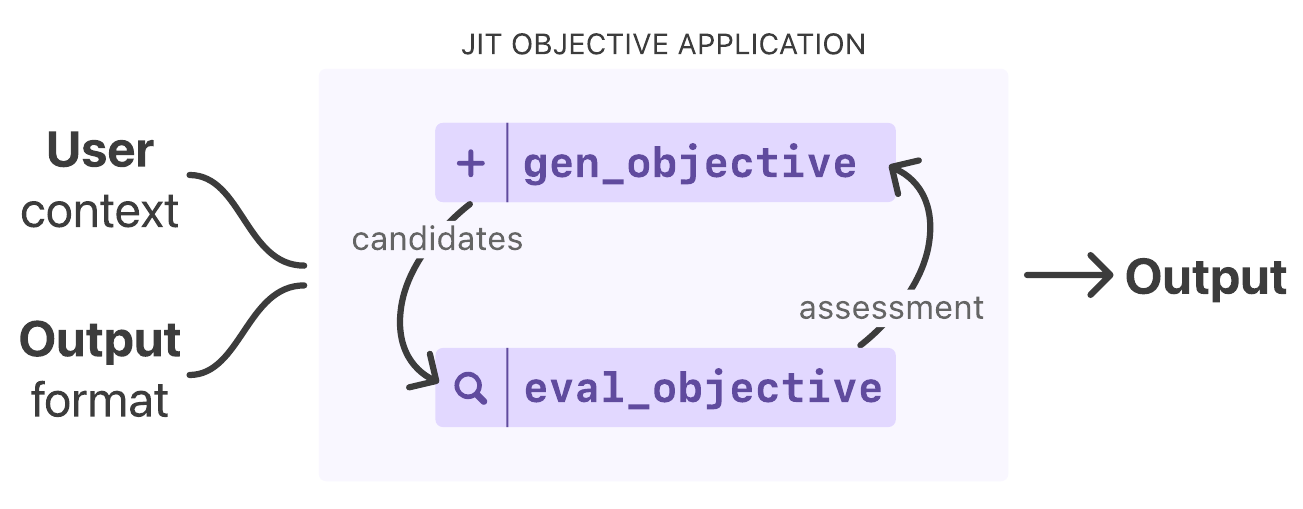}
    \caption{\textbf{Applying Just-In-Time Objectives}. The core of our architecture is a just-in-time objective that can add on to LLM system components, such as those responsible for generation and evaluation. Here, we show an example of dual generation (\texttt{gen\_objective}) and evaluation (\texttt{eval\_objective}) steps that optimize model behavior. The generation step produces candidates of the specified output format based on the user context and objective, and the evaluation step assesses these candidates with respect to the objective.
    }
    \label{fig:arch_loop}
    \Description{Architecture diagram showing how just-in-time objectives enhance generation and evaluation components in language model systems. A flowchart depicts the core architecture with three main components: (1) inputs of User context and Output format, (2) a core operation of Generation and Evaluation, and (3) Output resulting from this process.}
\end{figure*}

\subsubsection{Generate: Applying JIT Objectives to Optimize Generation}
The most common (and central) components of LLM systems are those responsible for \textit{generation} (\autoref{fig:arch_loop}). Generators are akin to actors in a traditional actor-critic architecture: they leverage the objective to make suggestions. We use the umbrella term \textit{generator} to refer to any language model-based system that takes in user input and produces a new artifact: a traditional LLM chat assistant is a typical example, but other examples include feedback and review generators, design and brainstorming tools, role-prompting pipelines, and coding assistants~\cite{xu2023expertprompting, wu2023autogen, fourney2024magentic, choi2025proxona, park2024thinkingassistant, ashkinaze2024plurals}. 

\begin{quote}
    \textit{Mary is writing the System section of her CHI paper, and she wants feedback to make it more compelling. She uploads the draft and asks an LLM to ``Give feedback on this draft,'' but receives generic suggestions about verbose sections to cut, technical jargon to simplify, and low-level implementation clarifications.}
\end{quote}

\begin{figure*}[!tb]
    \centering
    \includegraphics[width=0.65\linewidth]{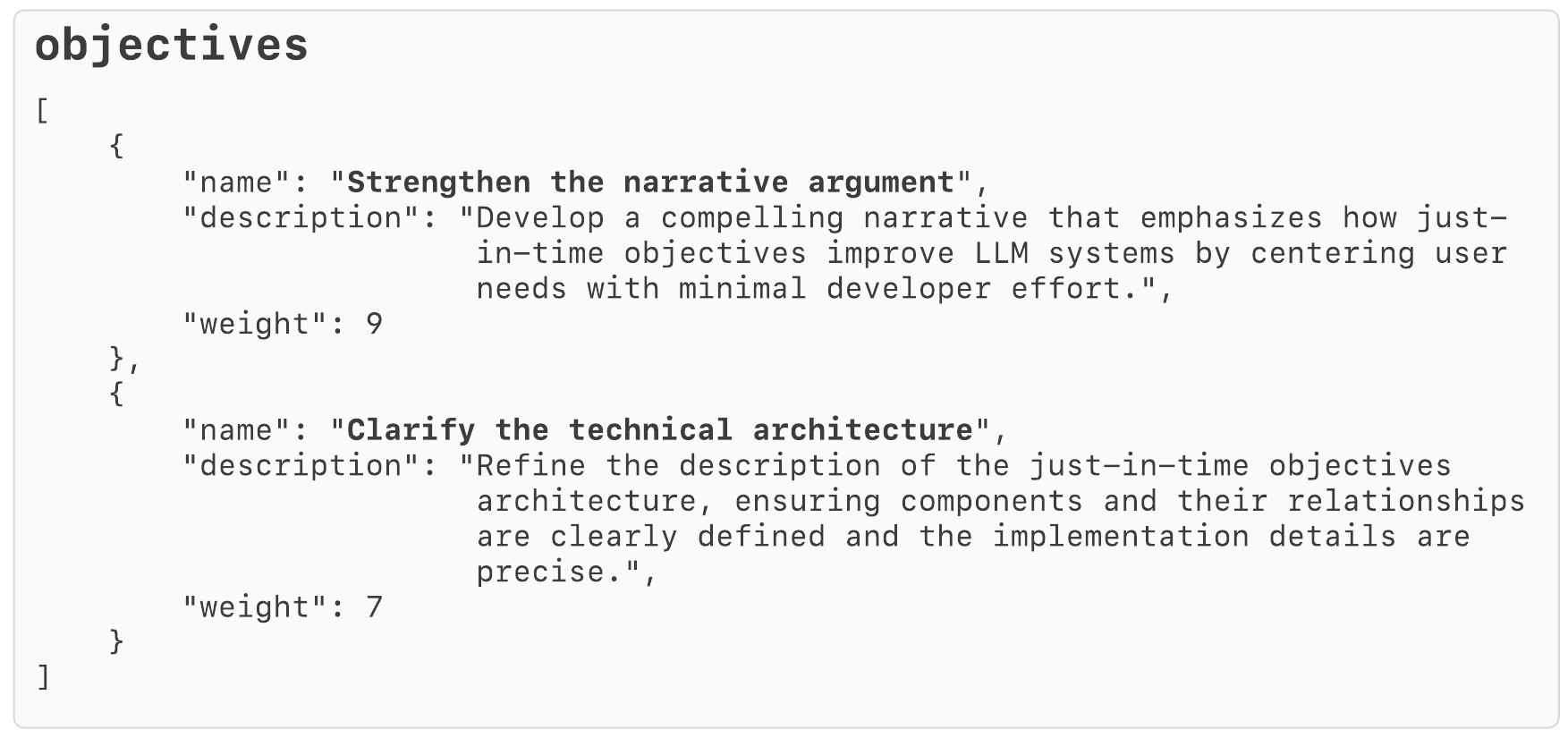}
    \caption{The JSON specification for just-in-time objectives includes a name, 1-2 sentence description, and weight indicating the estimated importance of the objective on a 1-10 scale.}
    \label{fig:objectives_spec}
    \Description{JSON specification format for just-in-time objectives containing name, description, and weight fields.}
\end{figure*}

Our \texttt{gen\_objective} operator applies a just-in-time objective to a provided generator to optimize its output towards a user's goals. 
We instantiate it as a small snippet of context (Figure~\ref{fig:objectives_spec}) prepended to existing instructions for generation.

\begin{quote}
    \textit{Mary enables objective induction based on editing patterns in her paper and appends the result to her existing prompt. Now, she gets more varied and interesting suggestions. The system induced an objective of ``Strengthen the narrative argument,'' which steered the system to provide advice on how to move beyond sequentially describing system components and instead present a narrative user walkthrough of the system.}
\end{quote}
The \texttt{gen\_objective} operator can apply to any generation call, so objectives can also be applied to generators that revise existing content, initiate tool calls, or perform actions on behalf of the user.

\subsubsection{Evaluate: Applying JIT Objectives to Optimize Evaluation}
Many LLM systems include an \textit{evaluation} component to assess whether generations are working as intended. Such components are akin to the critic in an actor-critic architecture. Just-in-time objectives can shape evaluators as well. We use the term \textit{evaluator} to refer to a language model-based system that takes in a generated artifact and produces an assessment or evaluation of that artifact. For example, this includes LLM-as-a-judge~\cite{zheng2023llmAsJudge} modules that take in model-generated text and evaluate its quality with respect to predefined rubric items to generate a rating or score. Other examples include systems that aid iteration on criteria, provide critique on model generations, or perform comparative sorting or ranking of provided model generations~\cite{kim2024evalLM, shankar2024evalGen}.

\begin{quote}
    \textit{To try to get better feedback, Mary tries a standard LLM-as-a-judge setup to evaluate feedback options. She creates a rubric that assesses the overall quality and intellectual rigor of the feedback to select for higher-level feedback on the content and ideas rather than presentation. Without a just-in-time objective, she finds that many of the feedback options receive similarly high scores, and do not differentiate themselves.}
\end{quote}

The \texttt{eval\_objective} operator applies a JIT objective to a provided evaluator to optimize its assessment towards a user's goals. This can again be implemented simply by adding the induced objective JSON specification (Figure~\ref{fig:objectives_spec}) to an existing evaluation prompt.

\begin{quote}
    \textit{Adding the JIT objective to her LLM-as-a-judge prompt, Mary gets a larger spread of scores. For example, scoring feedback on the induced JIT objective of ``Strengthen the narrative argument'' allowed the system to filter out feedback that merely mentioned the importance of narratives, but did not provide concrete strategies on how to incorporate a narrative into her draft.}
\end{quote}

Added on to a recommendation module that selects the highest-relevance candidates, \texttt{eval\_objective} can steer the system towards objective-aligned options. Added on to a critique module that produces feedback on generated candidates, \texttt{eval\_objective} can aid iterative refinement loops by producing objective-based critiques that feed back into a generator to improve its next round of candidates. Added on to a sampling module that selects the best of N generations, \texttt{eval\_objective} can serve as a verifier for test-time scaling towards user goals. 

\begin{figure*}[!tb]
    \centering
    \includegraphics[width=0.6\linewidth]{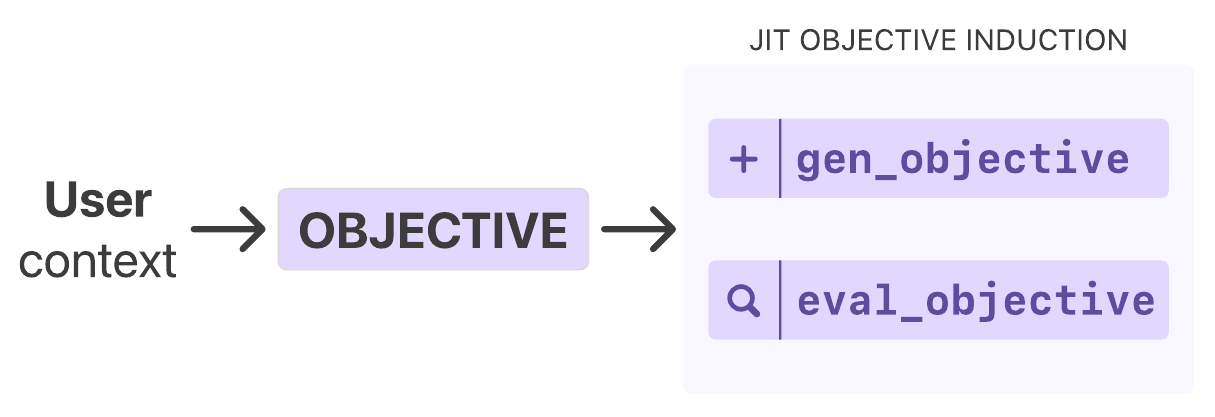}
    \caption{\textbf{Inducing Just-In-Time Objectives}. The core architecture depends on a single upstream step of \textit{objective induction} that infers user goal(s) based on observed context and instantiates each goal as a \textit{just-in-time objective}, which can later be applied to optimize generation (\texttt{gen\_objective}) and evaluation (\texttt{eval\_objective}). 
    }
    \label{fig:arch_induction}
    \Description{Architecture diagram showing how user context observation leads to an objective, which produces generation and evaluation operators that can steer downstream AI systems.}
\end{figure*}

\subsubsection{Inducing JIT Objectives}
We've described how \texttt{gen\_objective} and \texttt{eval\_objective} apply a JIT objective to generation and evaluation tasks, but how do we produce this objective in the first place? The objective induction step aims to infer likely user goals and translate them into just-in-time objectives. We define a \textit{goal} as a user's intention or aim for a window of time. Meanwhile, we define an \textit{objective} as a system's operationalization of the user goal into a form that can be used to guide optimization.

Objective induction takes in user context, infers likely user goals, and converts those goals into our just-in-time objectives, which can be subsequently applied to steer downstream AI systems (\autoref{fig:arch_induction}). An objective includes a brief name, a 1-2 sentence detailed description, and a weight that indicates the estimated importance of the objective on a 1-10 scale~(\autoref{fig:objectives_spec}).

We perform objective induction by taking in user input data in the form of text, images, or file attachments. This input user context may take the form of a text string (e.g., content from Overleaf or Google Docs), a workspace screenshot (e.g., the user's full browser window with a site like Figma, Google Slides, or Google Sheets, or their full desktop with an application like VSCode or Microsoft Word), or an attached file (e.g., an image of a poster draft, a PDF of a research article). We then prompt a vision language model (VLM) to induce objectives: we include the input user context, ask the system to infer goals, and request output in the form of the objective JSON specification~(\autoref{fig:objectives_spec}). Our objective induction prompt includes a chain-of-thought process that has the model reflect on factors including the user's task domain, stage of work completion, potential audience, ideal final task output, and anticipated reaction to assistance (full prompts in Appendix~\ref{appendix:prompts}).

We present this approach as a simple-to-implement yet effective architecture. One might improve on it by, for example, utilizing trajectories over time rather than a single cross-sectional snapshot. In addition, users can modify the objective induction prompt to steer just-in-time objectives at a meta level. For example, some users may want to adjust the context window of an objective to have the system only infer in-the-moment objectives for their actions in the next minute (e.g., ``Highlight and delete other usages of an outdated system name'') while others may want the system to propose longer-range objectives across many weeks (e.g., ``Improve the clarity of my academic writing''). 

\begin{figure*}[!tb]
    \centering
    \includegraphics[width=\linewidth]{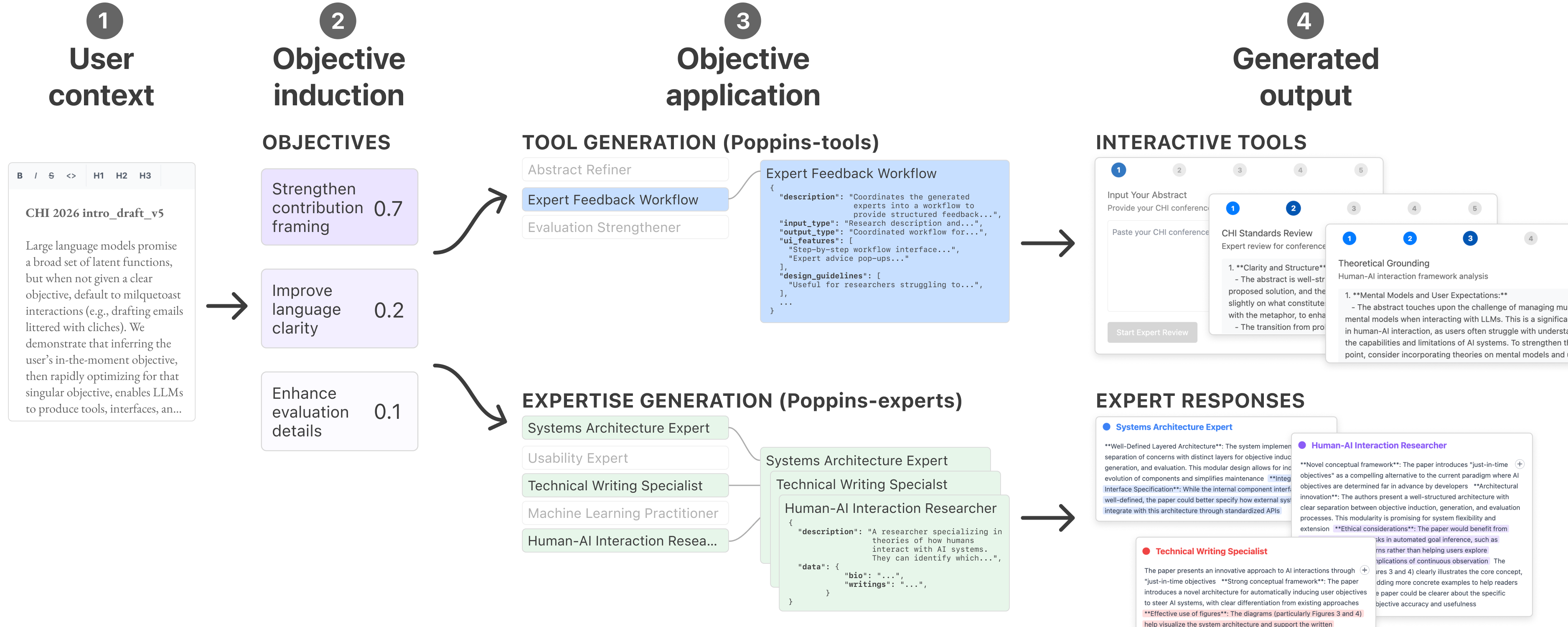}
    \caption{\textbf{\systemName{} incorporates just-in-time objectives to power UI generation}. The \systemName{} system instantiates our just-in-time objectives architecture by inducing objectives from a user's context (e.g., their screen or text document). Then, generation and evaluation steps steer objective application for \textit{Tool Generation} (Poppins-tools) and \textit{Expert Generation} (Poppins-experts) to automatically generate a tailored output for the user's task (interactive software tools or expert responses, respectively).}
    \label{fig:system}
    \Description{A system diagram showing how Poppins uses just-in-time objectives to power two main functions: Poppins-tools (interactive tool generation) and Poppins-experts (expert response generation). The architecture shows user context input flowing through objective induction, then through generation and evaluation steps guided by the objectives, ultimately producing tailored outputs for the user's specific task.}
\end{figure*}

In sum, the lightweight JIT objective architecture that can be implemented into any end-user interactive system is: (1)~leverage the user's current state to induce a just-in-time objective, (2)~apply that objective in context to shape any generated content, and (3)~apply that objective again to shape any automated evaluations of the generations. Below, we demonstrate how this architecture enables improved interaction in a web platform.

\subsection{Poppins: UI Generation Powered by Just-In-Time Objectives}
Having outlined how we construct just-in-time objectives, we demonstrate how system developers can apply the just-in-time objectives architecture to steer LLMs in interactive systems.
We develop \systemName{}, a system for UI generation that incorporates just-in-time objectives (\autoref{fig:system}). \systemName{} is a browser extension and web application that observes user screens and, when enabled, automatically generates outputs to assist the user in their specific task.\footnote{Much like our namesake Mary Poppins, whose bag always carried the right tool for the particular task at hand.} 
Critically, rather than requiring users to devote time and cognitive effort to craft requests for customized assistance, our system recognizes user needs and produces outputs that users might not have time to create, or might not have thought to create. Our system can take on this alternate task formulation only because it builds on just-in-time objectives that set automated generation and evaluation processes on the right user-aligned course. To demonstrate how JIT objectives can integrate into existing LLM systems of interest to the HCI community, \systemName{} supports two types of generated assistance: (1)~\textit{Expertise generation} with \systemName{}-experts and (2)~\textit{Tool generation} with \systemName{}-tools. These two types of assistance demonstrate how LLMs can better support common workflows with JIT objectives: (1) feedback on user input with expert perspectives~\citep[e.g.,][]{benharrak2024writerDefined, dArcy2024MARG, choi2025proxona}, and (2) on-demand interface construction and customization~\citep[e.g.,][]{cheng2024biscuit, vaithilingam2024dynavis, chen2025generativeinterfaces, cao2025jelly}.

\begin{figure*}[!tb]
    \centering
    \includegraphics[width=\linewidth]{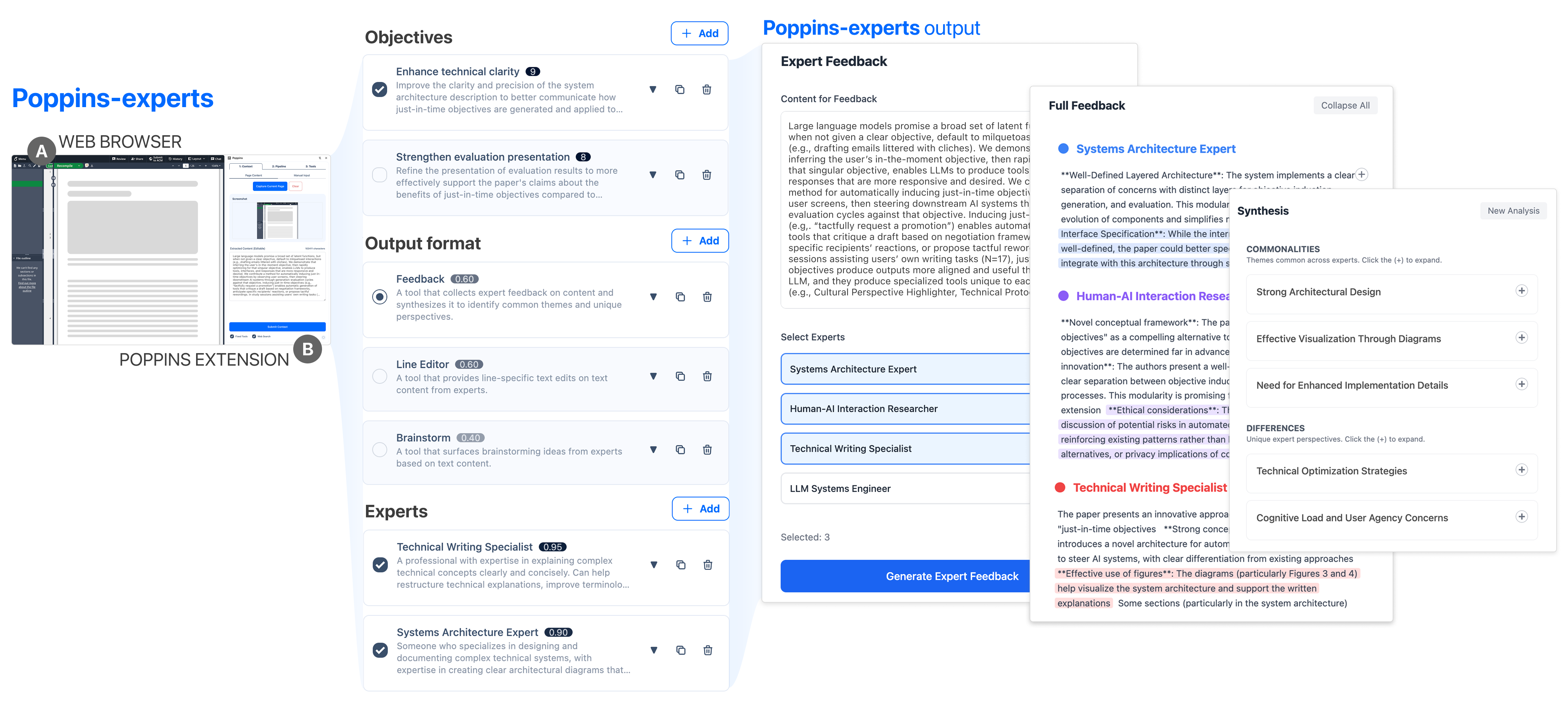}
    \caption{\textbf{Poppins-experts} uses just-in-time objectives to generate relevant model expertise, select an appropriate output format (e.g., Feedback, Brainstorm), and generate model responses drawing on the specified expertise. Users can view and modify all objectives, output formats, and experts.}
    \label{fig:poppins_experts}
    \Description{Poppins-experts interface displaying generated expert profiles and feedback outputs. A screenshot of the browser extension interface showing three generated expert profiles: Technical Writing Specialist, Systems Architecture Expert, and Human-AI Interaction Researcher. The interface shows the objectives, output format, and experts selected to create this output. The output interface shows input fields for the user's content and text responses from each expert, along with a synthesis of main commonalities and differences among the experts.}
\end{figure*}

\subsubsection{Input Pipeline}
As input, \systemName{} aims to capture as much of a user's current working environment, subject to user convenience and privacy preferences. The current implementation accepts image and text input, which allows the system to see what users are writing and reading, as well as a broad set of other visual content they may be creating or viewing on their device. The browser extension provides a one-click button to capture the currently-visible screen and all text content on the webpage, or users can manually upload their own files or copy their own text content. 

\begin{quote}
    \textit{Mary is working on a CHI paper draft in Overleaf, specifically the System section. She's been looking at it so long that she can't tell if it makes sense. Could \systemName{} help? She opens the extension sidebar and clicks the button to capture her window.
    Under the hood, based on the text being edited, the system infers candidate objectives of ``Enhance technical clarity'' (weight: 9) and ``Strengthen evaluation presentation'' (weight: 8) and selects the higher-weighted first option to translate into a just-in-time objective.
    }
\end{quote}

While these input modes provide a lightweight, non-invasive entrypoint into user workflows, our system is flexible to support other input sources, so it could be extended to support streams of periodic screen captures, continuous video recordings, webpage interaction logs, or revision histories for more rich and domain-specific context to inform objective induction.

\subsubsection{\textbf{Expertise Generation} with \systemName{}-experts}
\label{sec:system_poppins_experts}
The first form of assistance that \systemName{} implements is \textit{expertise generation}: it leverages the JIT objective to identify experts, perspectives, and areas of expertise that might be most helpful for assistance (\autoref{fig:poppins_experts}).

Using the JIT objective, \systemName{}'s expertise generator produces an expert specification that includes a name, description, and several paragraphs of detailed background material retrieved using LLM search, such as relevant publications, talks, and projects; specific expertise areas and methodologies; and key ideas or quotes to consider. 
\systemName{} appends a \texttt{gen\_objective} operator to steer expertise generation towards the user's objective. Then, it performs evaluation with an \texttt{eval\_objective} operator to score the candidate experts and select the top objective-aligned suggestions.

\begin{quote}
    \textit{Mary has \systemName{}-experts enabled, so the system applies the just-in-time objective to the expertise generator to surface background knowledge relevant to her paper and objective of ``enhancing technical clarity.'' This produces a ``Technical Writing Specialist,'' ``Systems Architecture Expert,'' and ``Human-AI Interaction Researcher,'' each of which is supported by detailed background information referencing particular academic papers, researchers, and concepts that bridge the expertise area and the content of Mary's paper (\autoref{fig:poppins_experts}). }
\end{quote}

After model expertise generation, \systemName{}-experts uses an output evaluator to determine the appropriate output format for this expert. While we could leave out this step and simply prompt the model using the expert specification, setting the desired output format affords greater control over expert output. We implement an initial set of output formats, each of which steers prompts to the experts and is mapped to a UI template to render the expert outputs: Feedback (generates feedback and a synthesis of common themes and unique perspectives), Brainstorm (generates brainstorming ideas based on input content), and Line Editor (interactively generates line-level edits on user-highlighted text). The output evaluator takes in the user context and selects the most relevant output format; \systemName{} adds on an \texttt{eval\_objective} operator to account for the user objective in this selection. 

\begin{quote}
    \textit{Based on Mary's Overleaf context and induced objective, \systemName{}-experts selects the Feedback output format and loads the associated Feedback UI in the extension sidebar with her paper content pre-loaded as input. Mary views the generated experts and clicks the ``Run'' button to proceed with gathering feedback. The resulting output surfaces helpful feedback from the Human-AI Interaction Researcher that the section ``could be clearer about the specific metrics used to assess objective accuracy and usefulness,'' and that the paper would benefit from ``deeper discussion of the cognitive load implications'' of users reviewing system's automatic inferences.
    Meanwhile, the System Architecture Expert raises that ``internal component interfaces are well-defined,'' but the paper ``could better specify how external systems would integrate with this architecture,'' which Mary flags as a potential topic in the Discussion.
    }
\end{quote}

Model expertise is not restricted to persona-like experts as shown above. The system can generate perspectives such as:
real-life individuals (e.g., specific HCI/AI researchers); communities (e.g., CHI subcommittees like Blending Interaction, Critical Computing); schools of thought (e.g., different HCI evaluation methodologies); fictional characters (e.g., \textit{Inside Out}'s emotion-based characters); abstract concepts (e.g., Human values from Schwartz's theory of Basic Human Values~\cite{schwartz2012refining}); and styles (e.g., the user's own professional versus humorous writing voices).

By default, Poppins-experts presents an automated workflow to both generate experts and expert response outputs. However, users can optionally modify any of the intermediate generations (Objectives, Experts, and Output formats) with several actions (\autoref{fig:poppins_experts}). They can \textit{Select} other options that did not receive the highest score, \textit{Edit} existing options by rewriting system-generated content in the form fields (e.g., expert name, description, background material), \textit{Add} new options and manually author all form fields, and \textit{Delete} any unwanted options.

\begin{figure*}[!tb]
    \centering
    \includegraphics[width=\linewidth]{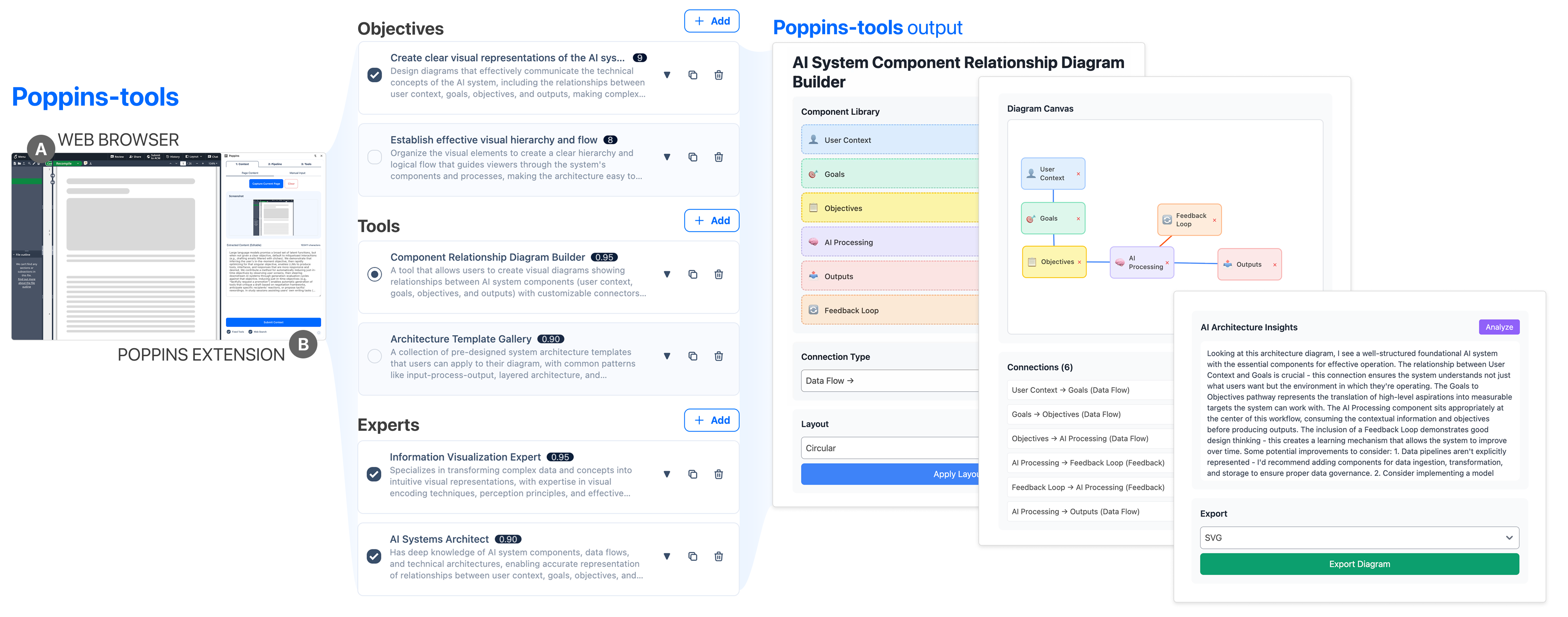}
    \caption{\textbf{Poppins-tools} applies just-in-time objectives to generate tool designs, optionally generate relevant model expertise, and generate UI code to instantiate the tool idea. Users can view and modify all objectives, tool designs, and (optional) experts. Here, the system induced an objective of ``Create clear visual representations of the AI system.'' Poppins-tools then generated a \textit{Component Relationship Diagram Builder}, a drag-and-drop diagramming UI to visualize system components, experiment with layouts, and request expert feedback on the system architecture diagram.}
    \label{fig:poppins_tools}
    \Description{Poppins-tools interface showing generated tool specifications and implementation. A screenshot displaying a Component Relationship Diagram Builder tool generated by the system. The interface shows the objectives, tools, and experts selected to create this tool. The implemented tool appears as a diagram canvas with draggable component blocks and connection options, along with an AI Architecture Insights button for expert feedback.}
\end{figure*}

\subsubsection{\textbf{Tool Generation} with \systemName{}-tools}
\label{sec:system_poppins_tools}
\systemName{} demonstrates a second form of assistance via \textit{tool generation}: based again on a user's context, \systemName{}-tools attempts to design and synthesize a code implementation for an interactive tool to assist the user (\autoref{fig:poppins_tools}).
Here, we define a ``tool'' as interactive software that assists the user, as opposed to static text responses (e.g., chat reply, feedback, draft document) that are typical of LLM chat systems and encapsulated by Poppins-experts. To demonstrate the space of Poppins-tools outputs, we include examples (Figures~\ref{fig:poppins_study_examples}, \ref{fig:poppins_tool_examples_2_study}, \ref{fig:poppins_tool_examples_3_team}).

We implement a tool design generator that describes a design idea for a tool: given the JIT objective, this generator produces a tool specification that includes a high-level description of its function (name and description fields), low-level descriptions of the implementation approach (input type, output type, descriptions of interface features, and descriptions of expected user behavior), and design guidelines outlining the problem settings for which this tool is an apt solution. \systemName{} adds a \texttt{gen\_objective} operator to the tool design generator to steer its tool ideas to meet user goals and an \texttt{eval\_objective} operator to score the tool ideas and select the most promising candidates.

\begin{quote}
    \textit{Mary found the expert feedback useful to aid her thinking, but she has now switched over to Figma to mock up a system architecture figure and wants more granular assistance.    
    She tries enabling \systemName{}-tools and initiates the system again, and this time it infers an objective of ``Create clear visual representations of the AI system.'' This objective produces tool suggestions of ``Component Relationship Diagram Builder'' (drag-and-drop tool to visualize connections between components in a system architecture diagram), ``Architecture Template Gallery'' (gallery of system architecture templates to apply to a diagram), and ``Component Style Synchronizer'' (tool to unify styles, colors and formatting across a system architecture diagram). The tool candidates are accompanied by interaction and interface design details (\autoref{fig:poppins_tools}).
    }
\end{quote}

After tool design generation, \systemName{}-tools runs a UI code generator that takes in the tool specification and optional expert specifications and generates a code implementation of the tool idea. This generator formulates a request for a standalone web interface implemented as a Svelte component (see prompts in Appendix~\ref{appendix:prompts}), and \systemName{} adds a \texttt{gen\_objective} operator to encourage objective alignment.
To provide more reliable model behavior around LLM integration, we provide the system with access to a library of LLM helper functions (with functions like \texttt{getExperts}, \texttt{promptExpert}, \texttt{promptGeneral}). These allow the system to more easily formulate LLM calls, especially to consistently incorporate the requested expertise if useful to the induced tool. 
Given the high cost of generating candidates for the UI generation task using more advanced models, the UI code evaluator is a critique module that provides feedback to refine the current candidate rather than selecting over a pool of candidates. The evaluator performs checks for usability, design quality, and bug-free code, and \systemName{} adds an \texttt{eval\_objective} operator to vet objective alignment. This step produces a final code implementation, which we render in \systemName{}.

\begin{quote}
    \textit{The implemented ``Component Relationship Diagram Builder'' now appears in Mary's extension sidebar, which created a library of components based on her system (e.g., User context, Goals, Objectives) that she can add to a diagram canvas (\autoref{fig:poppins_tools}). Here, she can drag and reposition component blocks and add connections of different types (e.g., data flow, feedback, dependency) between blocks. At the bottom, an ``AI Architecture Insights'' button retrieves feedback from the ``AI Systems Architect'' on the current diagram. 
    Mary explores several layouts in the diagram canvas before requesting feedback, and the ``AI Systems Architect'' indicates confusion between ``goals'' and ``objectives.'' Mary realizes that her current diagram focuses on laying out the terms aesthetically, but doesn't make clear what underlying data and operations happen within these steps, so she gets the idea to place a worked example below the component blocks in her figure.}
\end{quote}

Similar to Poppins-experts, Poppins-tools by default automatically generates tool designs and the tool UI code. Users can optionally modify intermediate generations (Objectives, Tools, and Experts) with the same actions: \textit{Select} other options, \textit{Edit} an existing option (e.g., rewriting the tool input type, output type, description), \textit{Add} a new option, or \textit{Delete} an unwanted option (\autoref{fig:poppins_tools}).

\subsubsection{System Implementation}
The web application is implemented with a Python Flask backend server and a Svelte frontend. The browser extension is implemented as a Google Chrome extension using Svelte components. The frontend server is hosted on Vercel, and the backend server is hosted with Heroku.
We use three LLMs in our system for differing purposes. 
By default, we use Claude Sonnet 3.7 as the core model for objective induction and generators in our system, as it performed reliably and cost-effectively for these functions. 
For our UI code generator, we use Claude Sonnet 4, as the earlier model was not capable of producing high-quality, consistent UI code implementations. 
Then, for our evaluators, we use GPT-4o mini, as our score-based evaluation is a straightforward classification task well-served by a smaller model, and it is useful to perform evaluation with an independent model provider relative to our generator. We use GPT-4o mini Search Preview for web search to populate detailed background information for the expert specifications given its speed and cost efficiency.

\subsection{System Limitations}
We note several limitations of the just-in-time objectives architecture and \systemName{} system. First, a just-in-time approach necessarily incurs a time cost since users must wait for the objectives to be induced and applied, which currently takes 1-3 minutes. Since this is much slower than a standard LLM chat response, JIT objectives may be appropriate only when the time investment is worthwhile for a higher quality response. 
Our method is heavily reliant on the model to accurately infer and apply user objectives. While our evaluation finds that objectives are highly accurate, LLMs may have inductive biases such that they are less likely to induce certain objectives, or less performant when applying certain objectives, warranting further evaluation before widespread release. 

The architecture has minimal visibility into a user via an individual snapshot of context, which can result in some inaccurate induced objectives. While this lightweight approach demonstrates a ``floor'' of performance, a single screenshot has limited visibility into complex tasks that require more context over time, or subtle needs that require more observations to understand.
While we can gather more data, there are limits to this strategy: there is no way to instrument a user to capture their full life context, and gathering more data can dramatically increase privacy and security risks for users. Longer context windows increase coverage of the necessary information to accurately infer objectives, but also introduce new challenges of effectively filtering out noise and handling contradictory evidence.

\systemName{} instantiates experts seeded with retrieved background knowledge, but these experts are not guaranteed to behave exactly as corresponding area experts would. While UI code generation is consistently bug-free for primary functionality, some implementations include minor bugs such as non-functional buttons or flawed LLM response parsing. These issues can be improved with further rounds of critique and by restricting the generator to vetted component libraries for settings where UI fidelity is important. While experts and tools can provide highly specific assistance, these scaffolds directly trade off against the flexibility of natural language, which some users may prefer.

\section{EVALUATION: Assessing the Accuracy and Utility of Just-In-Time Objectives}
\label{sec:survey_eval}

Our evaluation has two main aims: (RQ1) to determine whether our just-in-time objectives are accurate and useful, and~(RQ2) to investigate whether systems that incorporate just-in-time objectives are relevant and useful to users' needs.
To assess RQ1, we conduct experiments where participants evaluate just-in-time objectives induced from submitted screenshots of their own browsing behavior (Section~\ref{sec:survey_eval}). 
We assess RQ2 by conducting 1-hour in-lab sessions where participants receive assistance on their own writing task with a baseline LLM and the \systemName{} system that incorporates just-in-time objectives (Section~\ref{sec:poppins_eval}).
To establish a ``floor'' of performance, all evaluations used a single screenshot, the minimal context window, for objective induction. This evaluation setup allows us to assess JIT objective performance on a minimally-invasive setting of individual user screens.

In this section, we conduct a series of experiments to evaluate the accuracy and utility of just-in-time objectives and their resulting output generated from participants' realistic task environments.

\subsection{Procedure}
First, we run an evaluation on participants' own web browser traces ($N=\screenSurveyN{}$). The goal of this evaluation is to assess the \textit{accuracy and utility} of just-in-time objectives for a realistic set of tasks that participants encounter in their daily lives. Then, we expand our evaluation to assess how our system performs for a broader range of users and tasks ($N=\prolificStudyN{}$).

\subsubsection{Study 1: Assessing Accuracy \& Utility}
\label{sec:eval-screen-acc}
For the first study, we gather screenshots from participants' daily tasks as input contexts to induce just-in-time objectives. 
We recruit participants from university mailing lists and social media in accordance with our institution's IRB. Participants first submit a pre-survey with background information on their LLM use, along with a link to a document they will use for their writing task during the in-lab study session (Section~\ref{sec:poppins_eval}).
In total, \userStudyN{} participants took part in our study (see Appendix~\ref{appendix:demographics} for participant background information). Participants were compensated with a \$35 Amazon gift card for completing the hour-long lab study and an additional \$15 for participating in an accompanying screenshot study.

\begin{figure*}[!tb]
    \centering
    \includegraphics[width=\linewidth]{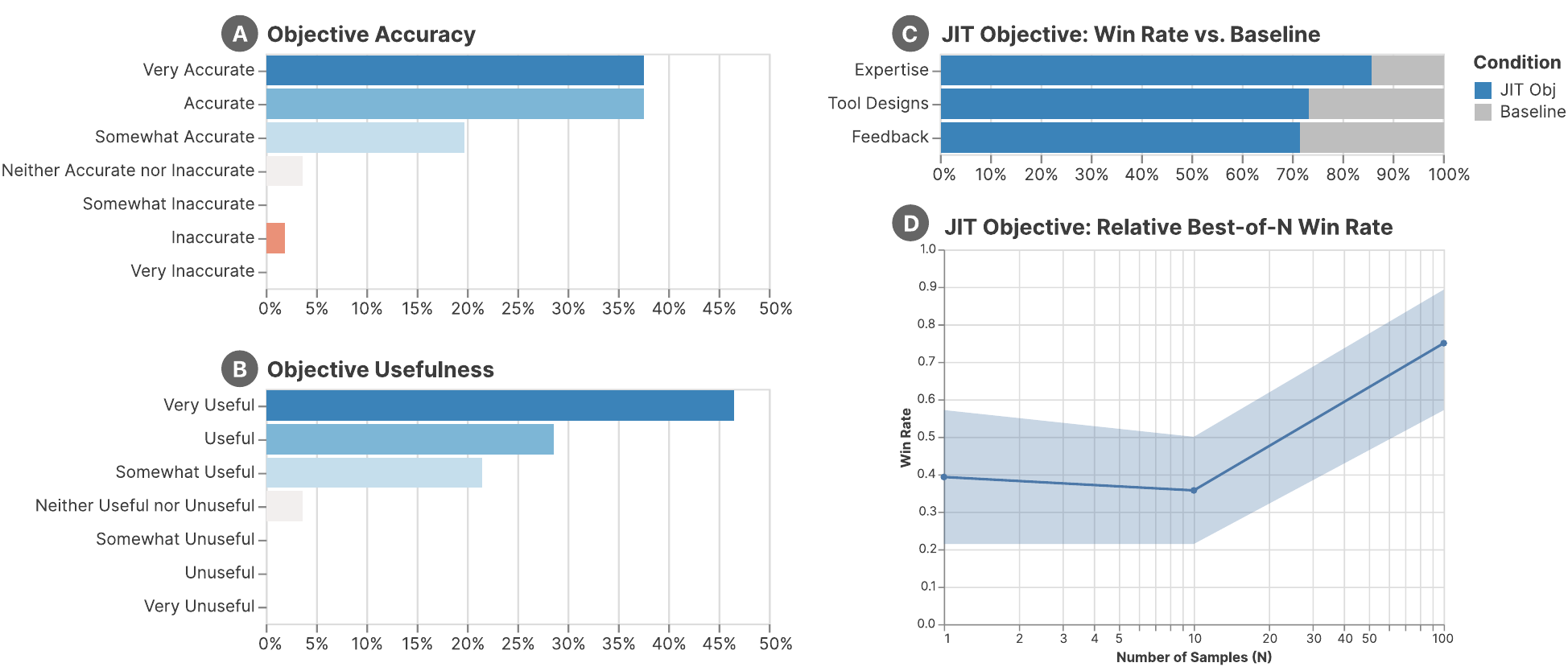}
    \caption{\textbf{Study 1: Induced from participants' own input contexts (70 inputs from $N=\screenSurveyN{}$ participants), just-in-time objectives are accurate, useful, and produce outputs preferred to a baseline LLM.} The vast majority of participants rate the induced objective as accurate and useful (A, B). Experts, tool designs, and feedback generated with a JIT objective are preferred over that of a baseline LLM in 71--86\% of contexts (C), and sampling over JIT generations with a JIT evaluator produces preferred outputs (D).}
    \label{fig:screen_survey}
    \Description{Survey results from 14 participants showing just-in-time objectives are accurate, useful, and produce preferred outputs. Four bar charts showing evaluation results: (A) Objective accuracy ratings on a 7-point Likert scale with 75\% rated as accurate or very accurate, (B) Objective usefulness ratings with 75\% rated as useful or very useful, (C) Win rates for just-in-time objectives versus baseline large language model across experts (86\%), tool designs (73\%), and feedback (71\%), and (D) Best-of-N evaluation showing increasing preference for higher sample sizes with just-in-time evaluator.}
\end{figure*}

We asked participants to complete a screenshot data collection task that is used for an optional study, and \screenSurveyN{} of \userStudyN{} participants completed this additional study.
To gather screenshot data, participants install a lightweight Chrome extension that allows them to capture their browser (a screenshot of the visible window and extracted text content from the page) and submit the data to our study. We instructed participants to share any browser content that was representative of their daily tasks and work. All surveys and instructions are included in Appendix~\ref{appendix:surveys}. We collect browser data over the course of three days, resulting in an average of 11 screenshots per participant. 
The submitted browser data spanned a range of interfaces (e.g., Google Docs, Google Slides, LLM chats, and academic paper databases), task types (e.g., presentation feedback, writing feedback, text comprehension, brainstorming), and topic areas (e.g., chemistry, neuroscience, natural language processing, linguistics, design, energy policy, TA training). 

For each participant, we run a data selection script on their submitted browser data to select five items as inputs to induce just-in-time objectives, providing a total of $n=70$ input contexts across participants. The script enforces criteria about (1)~\textit{suitability}: whether the provided item would benefit from AI assistance, (2)~\textit{quality}: whether the screenshot has enough context to understand the user's task, and (3)~\textit{divergence}: whether the item sufficiently diverges from other suitable items, to encourage a spread of differing tasks for each user (e.g., so that a user's tasks are not all academic writing). Thus, the script allows us to focus our study session around a set of informative real-life moments at which our system might intervene, while providing some consistency across participants. The selected images are used to generate a 20-minute study with two sections:

\begin{itemize}
    \item \textit{Objectives and generator task}. The first section allows us to assess the quality of our just-in-time objectives and JIT generator. For four of the five selected inputs, participants assess the usefulness and accuracy of the highest-weighted induced objective based on the input. Then, they provide pairwise preferences for experts, tool designs, and feedback generated using a just-in-time objective versus a baseline. This baseline condition received the same user screenshot, prompt, and LLM (Claude Sonnet 3.7) as the JIT objective condition, but just did not include an induced objective.
    \item \textit{Evaluator best-of-N task}. The second section allows us to investigate the efficacy of our JIT evaluator added on to a JIT generator. A common practice for producing high quality output from an LLM involves sampling many ($N$) candidate outputs and selecting the best (best-of-$N$)~\cite{stiennon2020learning, snell2024scaling, beirami2024theoretical}. A strong JIT evaluator should produce better outputs as $N$ increases. The intuition here is that a model gets more ``attempts'' to produce something strong, with the JIT evaluator returning the best final candidate. To evaluate this, participants provide pairwise preferences on output evaluated with a just-in-time objective, with progressively increasing sample sizes. For each sample size ($N=1, 10, 100$), we generate $N$ candidates using a JIT generator, and then select a single best candidate using a JIT evaluator. The best candidate from each sample is presented to the user (the ``best-of-N'' output).
\end{itemize}

\begin{figure*}[!tb]
    \centering
    \includegraphics[width=\linewidth]{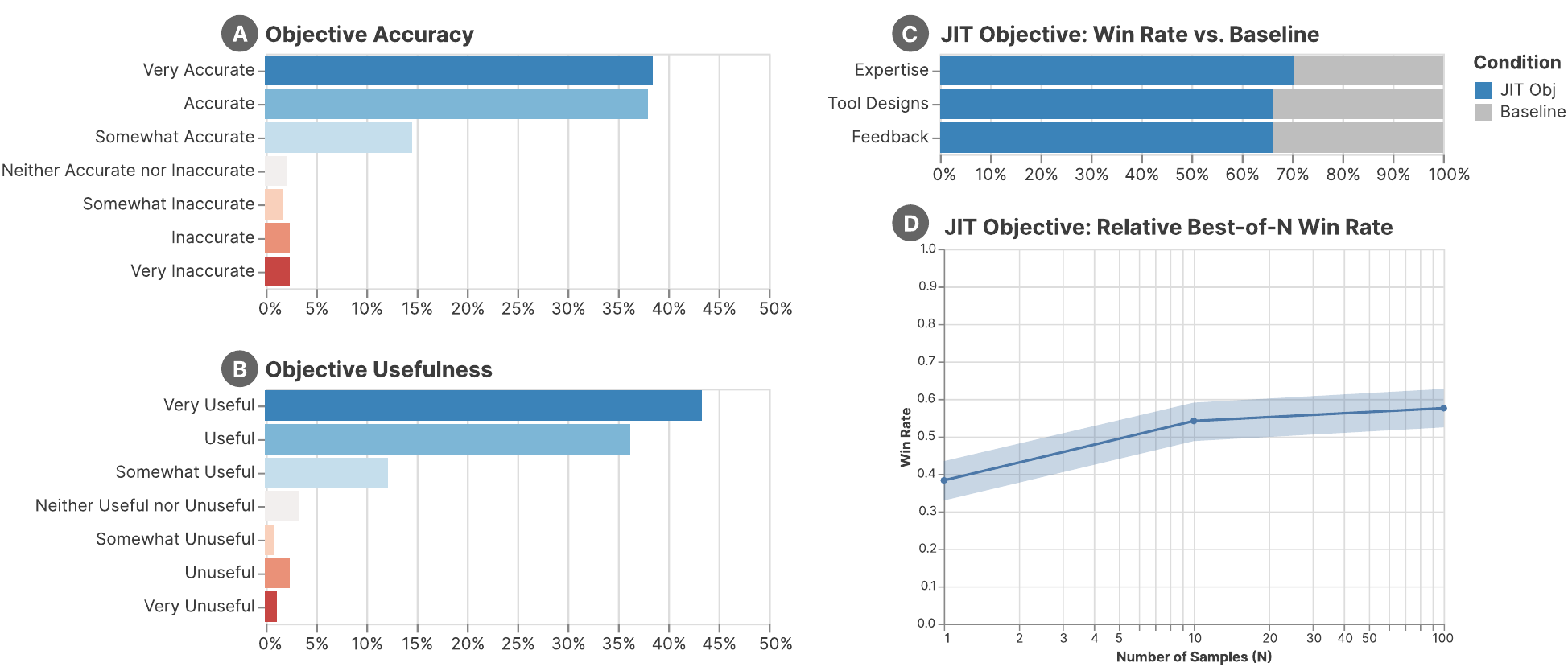}
    \caption{\textbf{Study 2: Across an expanded set of participant-submitted input contexts (410 inputs from $N=\prolificStudyN{}$ participants), JIT objectives are accurate, useful, and produce outputs preferred to a baseline LLM.} The vast majority of participants rate the induced objective as accurate and useful (A, B). Experts, tool designs, and feedback generated with a just-in-time objective are preferred over that of a baseline LLM (C), and sampling over JIT generations with a JIT evaluator produces more preferred outputs, though with marginal benefit beyond 10 samples (D).}
    \label{fig:prolific_survey}
    \Description{Expanded survey results from 205 participants confirming just-in-time objective effectiveness across broader contexts. Four bar charts mirroring Figure 8 structure but with larger sample size: (A) 77\% of objectives rated accurate or very accurate, (B) 80\% rated useful or very useful, (C) Win rates of 71\%, 66\%, and 66\% for experts, tool designs, and feedback respectively, and (D) Best-of-N results showing diminishing returns beyond 10 samples.}
\end{figure*}

\subsubsection{Study 2: Exploring Generalizability}
In the second evaluation, we conduct a study with an expanded set of online participants ($N=\prolificStudyN{}$) who provide screenshots of their workspace and evaluate resulting system-generated outputs.
The online study enrolls participants recruited from Prolific. At the start of the study, participants are asked to upload two screenshots of their desktop that capture a moment in time when they might benefit from AI assistance. These inputs are submitted to our system to perform live objective induction. Then, participants are asked to rate the usefulness and accuracy of the highest-weighted induced objective for their two task screenshots along with the most important goal for this task. Here, they can select from the three top-weighted induced objectives (presented in randomized order) or enter their own custom goal in a text field. The system then generates experts, tool designs, and feedback for the tasks based on the selected goal for each task.
The study then takes the same form as the prior study: participants complete a \textit{Generator} task and an \textit{Evaluator} task based on their uploaded screenshots. Participants were compensated through Prolific at a \$16/hour rate.
The study produced $n=410$ participant submissions, which spanned an even broader range of interfaces (e.g., word processors, spreadsheet software, presentation software, LLM chats, image/audio/video editing tools, design tools, code editors) and task types (e.g., creative writing, academic writing, travel planning, financial planning, health and fitness planning, resume drafting, letter writing).

\subsection{Results}
Across the two evaluation settings, we find that our induced objectives are rated as highly useful and accurate, and outputs produced with the aid of our just-in-time objective (experts, tool designs, and feedback) perform strongly above those of a baseline unsteered LLM~(\autoref{fig:screen_survey} and \autoref{fig:prolific_survey}).  

\subsubsection{Just-in-time Objective Accuracy and Usefulness}
\label{study1_jit_obj}
For both studies, we observe strong ratings on the accuracy and usefulness of induced objectives. On our 7-point Likert scale from -3: Very Inaccurate to 3: Very Accurate, we observe high accuracy ratings for Study 1 ($M=2.04$, $SD=0.5$) and Study 2 ($M=1.92$, $SD=1.07$), with a strong majority of objectives rated as ``Accurate'' or ``Very Accurate'' (Study 1: 75.0\%, Study 2: 76.6\%) (\autoref{fig:screen_survey}A, \autoref{fig:prolific_survey}A).
Similarly, with a 7-point Likert scale from -3: Very Unuseful to 3: Very Useful, we observe strong usefulness ratings for Study 1 ($M=2.18$, $SD=0.55$) and Study 2 ($M=2.06$, $SD=0.98$), and the vast majority of objectives were rated as ``Useful'' or ``Very Useful'' (Study 1: 75.0\%, Study 2: 79.8\%) (\autoref{fig:screen_survey}B, \autoref{fig:prolific_survey}B).

\begin{figure*}[!tb]
    \centering
    \includegraphics[width=0.5\linewidth]{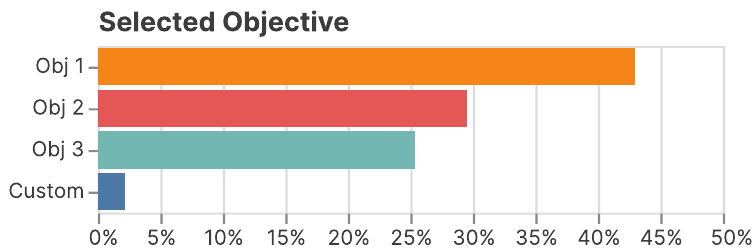}
    \caption{\textbf{Participants' selected objectives are aligned with just-in-time objective inferred weights ($N=\prolificStudyN{}$)}. Objectives 1-3 were generated with a JIT objective, where Obj 1 is the highest-weighted candidate, followed by Obj 2, then Obj 3. Objective options were presented in randomized order, and participants could enter a Custom objective if desired.}
    \label{fig:prolific_selected_goal}
    \Description{Participant objective selection aligned with system-inferred weights across 205 participants. A bar chart showing participant selection frequency for different objective options. Objective 1 (highest weighted) was selected 43\% of the time, Objective 2 selected 30\%, Objective 3 selected 25\%, and custom objectives only 2\%.
}
\end{figure*}

For Study 2, we have additional insight into which goal users deemed most important for their tasks. We find that just-in-time objectives were chosen in the vast majority of cases (97.8\%) over a custom-authored objective~(\autoref{fig:prolific_selected_goal}). These results cannot be interpreted as 97.8\% accuracy, as authoring a custom objective is more effortful than selecting an existing objective. However, this finding indicates that induced objectives were sufficiently relevant such that almost all participants did not opt to write their own objective.
In addition, participants' selections are well-aligned with the inferred weights associated with induced objectives: when chosen from a randomized ordering, the highest-weighted objective (Obj 1) was selected as the most important in 42.9\% of cases, followed by Obj 2 in 29.5\% and Obj 3 in 25.4\%.

\subsubsection{Just-in-time Objectives for Generation}
\label{study1_jit_gen}
Then, comparing the actual LLM outputs, we observe a consistent participant preference for those produced with the aid of just-in-time objectives over those produced with the baseline LLM.
In Study 1, just-in-time objectives achieved strong win rates for expertise (85.7\%), tool designs (73.2\%), and feedback (71.4\%) produced in response to participant screenshots (\autoref{fig:screen_survey}C). In Study 2, just-in-time objectives again won consistently against the baseline LLM for expertise (70.5\%), tool designs (66.3\%), and feedback (66.1\%) (\autoref{fig:prolific_survey}C).

Qualitatively, the baseline condition produces results that are relevant to a user's \textit{context}, but JIT objectives produce results that advance a user's specific \textit{goals}. For example, one participant submitted a screenshot of TA training handbook draft that had an unresolved comment about clarifying the role of TAs. The baseline produced outputs that are relevant to a TA handbook, but do not address this action item: a tool for helping TAs understand different course options, feedback on teaching strategies to add to the document, and expertise from a first-time TA. However, because the induced JIT objective was ``Enhance TA role description,'' the JIT generator produced outputs that were much more useful to the participant: a tool to expand descriptions of TA responsibilities with concrete examples of professional behavior, feedback on specific examples that could be added to the TA role, and expertise from an experienced TA.

\subsubsection{Just-in-time Objectives for Evaluation}
\label{study1_jit_eval}
Our best-of-N evaluations explore whether JIT objectives can serve as effective evaluators. In both studies, we find that JIT objectives allow us to sample effectively to produce preferred candidates even from among a pool of JIT generations: Best-of-100 outputs achieve a win rate of 75.0\% in Study 1 and 57.6\% in Study 2 over Best-of-10 and Best-of-1 (i.e., no evaluation) outputs (\autoref{fig:screen_survey}D, \autoref{fig:prolific_survey}D). 
Across head-to-head comparisons, the higher-N candidate won at higher rates (Study 1: 61.9\%, Study 2: 59.2\%), indicating that the JIT evaluator helps to sift through the noise and select more user-aligned candidates.
However, there may be diminishing returns for additional sampling: in Study 2, Best-of-100 only achieved a 3.5\% improvement in win rate. One reason may be that it is already possible to max out performance at a small sample size for these tasks, so additional samples merely reproduce the same caliber of results. Another possibility is that to better support test-time scaling using the JIT evaluator, we may need to introduce greater variation in our JIT generations.

Finally, though this study was primarily designed as a rating task, participants proactively shared their excitement about the JIT outputs in an optional final text field of the study, perhaps compelled because the system surfaced AI feedback on tasks from their own daily lives.
Participants indicated that the system surfaced valuable ideas that hadn't occurred to them before, such as P27: ``\textit{These were a series of short stories and it never occurred to me to link them all together in one story with a cohesive narrative, I'm thrilled about it!}'' and P56: ``\textit{The feedback for the code is fantastic. I didn't realize that I had vulnerabilities that could be exploited.}'' Some participants indicated surprise at what the system could do with a screenshot alone: ``\textit{I'm very impressed at how this AI bot understands what I need to do}'' (P41), ``\textit{I'm surprised how helpful these responses are, honestly}'' (P123). These participant responses present promising evidence that just-in-time objectives can unlock utility to users beyond what they typically encounter with available LLMs.

\subsection{Error Analysis \& Limitations}
Investigating the generated outputs, we find that experts generated using a JIT objective offered greater task-specific depth than baseline experts.
Tool designs generated with JIT objectives succeeded because they tended to facilitate a user's specific goals rather than context-related actions.
However, this specificity could be a double-edged sword: when users were in the early stages of work (e.g., brainstorming) or performing straightforward tasks (e.g., text comprehension), they often favored baseline tool designs. In such cases, users preferred lower-level, low-effort aids (e.g., ``character voice generator'') that the baseline tended to produce rather than higher-level, meta-cognitive tools (e.g., ``character development framework'') that a JIT objective would often produce. Feedback, by contrast, showed the most variable performance: we found no consistent relationship between task type or domain and preference, suggesting that feedback preferences may be more user-dependent.  

Our JIT \textit{generator} evaluation (Section~\ref{study1_jit_gen}) runs a comparison to assess whether JIT generators improve upon baseline LLMs. Meanwhile, the JIT \textit{evaluator} evaluation (Section~\ref{study1_jit_eval}) is designed to assess whether there is additional benefit in running a JIT evaluator on top of a JIT generator to verify that outputs match an induced objective. This evaluation design does not compare a JIT evaluator against a generic LLM-as-a-judge evaluator (i.e., one that selects ``the best candidate'' with no knowledge of the JIT objective). 
However, we qualitatively observe that JIT-selected outputs were successful by strongly encoding the \textit{objective}, which was \textit{not necessarily} aligned with a generic notion of quality. 

To further investigate this difference, we additionally ran a follow-up analysis on Study 2 using a generic LLM-as-a-judge prompt without the JIT objective. In cases where the higher-N result was preferred (in other words, where the user validated that the JIT evaluator had selected a better candidate), a performant evaluator should achieve strong agreement. We find that the generic evaluator prompt achieved only moderate agreement with the JIT evaluator on these cases ($\kappa$ = 0.46) and overall ($\kappa$ = 0.34), indicating that the generic evaluator displays different selection behavior and would not perform equivalently to the JIT evaluator.

A limitation of this study design is that participants chose their own screenshots to provide, which may have been biased towards certain kinds of tasks where they anticipated AI assistance may be helpful. We opted for this design to understand participant preferences on tasks that mattered to them, but future studies should experiment with eliciting a broader set of input contexts from users.

\section{EVALUATION: Powering UI Generation with Just-In-Time Objectives}
\label{sec:poppins_eval}

Our in-lab sessions explore whether JIT objectives allow our \systemName{} system to produce tools that are useful and relevant to users. 

\subsection{Procedure}
We conduct 1-hour sessions with participants ($N=\userStudyN{}$) where they apply provided tools to their own writing task. 
Participants experiment with both using these generated tools and modifying them to refine their behavior. 
This study was designed for qualitative feedback and was not intended or powered for statistical comparisons across conditions.

\subsubsection{Protocol}
Our recruitment process is described in Section~\ref{sec:eval-screen-acc}. We evaluate system-generated and baseline outputs in a 1-hour session with each participant according to the following structure:

\noindent\textit{1: Starter Phase (5 min)}.
Participants begin the session with consent and a brief study overview. They are instructed to think aloud throughout the duration of the session.

\noindent\textit{2: Baseline Comparison Phase (30 min)}.
The first phase compares tools that incorporate just-in-time objectives (\systemName{}) with those that do not (Baseline).
Condition order in this comparison phase is randomized across participants (\textit{Baseline-Poppins-Interview} or \textit{Poppins-Baseline-Interview}) as a counterbalancing measure for potential learning effects. 

\begin{itemize}
    \item \textit{Baseline}. In this phase, the participant uses a baseline system embedded within our extension that provides a standard LLM chat-style entry point to model outputs. This interface includes an input field for content or instructions and a text field that displays the model response, and is served by Claude Sonnet 3.7.
    \begin{itemize}
        \item Generate and review: Participant requests assistance from the baseline tool and reviews the output.
        \item Iterate: Participant iterates on chat responses to better align LLM outputs with their needs.
        \item Rate: Participant completes a brief survey rating their experience with this system variant.
    \end{itemize}
    \item \textit{\systemName{}-experts}. In this phase, the participant uses \systemName{}-experts, which performs objective induction and JIT generation and evaluation to generate expert responses. As described in Section~\ref{sec:system_poppins_experts}, Poppins-experts selects from a pre-implemented set of output formats, which provides greater consistency to facilitate comparison across participants.    
    \begin{itemize}
        \item System tutorial: Participant receives a short tutorial of the \systemName{} system features.
        \item Generate and review: Participant requests expert responses for their writing task context and reviews the output.
        \item Iterate: Participant iterates on expert generation with Poppins-experts to better align expert responses with their needs. Here, participants could modify any intermediate representations in the system (objectives, experts, output formats) to re-generate expert responses.
        \item Rate: Participant completes a brief survey rating their experience with this system variant.
    \end{itemize}
    \item \textit{Interview}. We conclude this phase with an interview to understand the participant's impressions of both conditions.
\end{itemize}

\noindent\textit{3: Exploratory Phase (20 min)}.
We conduct another task with \systemName{}-tools as a design probe. For this phase, we explore whether JIT objectives can perform the full UI generation task without intervention and how participants respond to this design concept.

\begin{itemize}
    \item \textit{\systemName{}-tools (10 min)}. Participant uses \systemName{}-tools to create an interactive tool based on their input context.
    \begin{itemize}
        \item Generate and review: Participant requests a tool for their writing task context and reviews the output.
        \item Rate: Participant completes a brief survey rating their experience with this system variant.
    \end{itemize}
    \item \textit{Interview}. We conclude with an interview to understand the participant's holistic experience with \systemName{} and the baseline, as well as gather feedback on how \systemName{} may fit into their everyday computer use.
\end{itemize}

\subsection{Results }

\begin{figure*}[!tb]
    \centering
    \includegraphics[width=\linewidth]{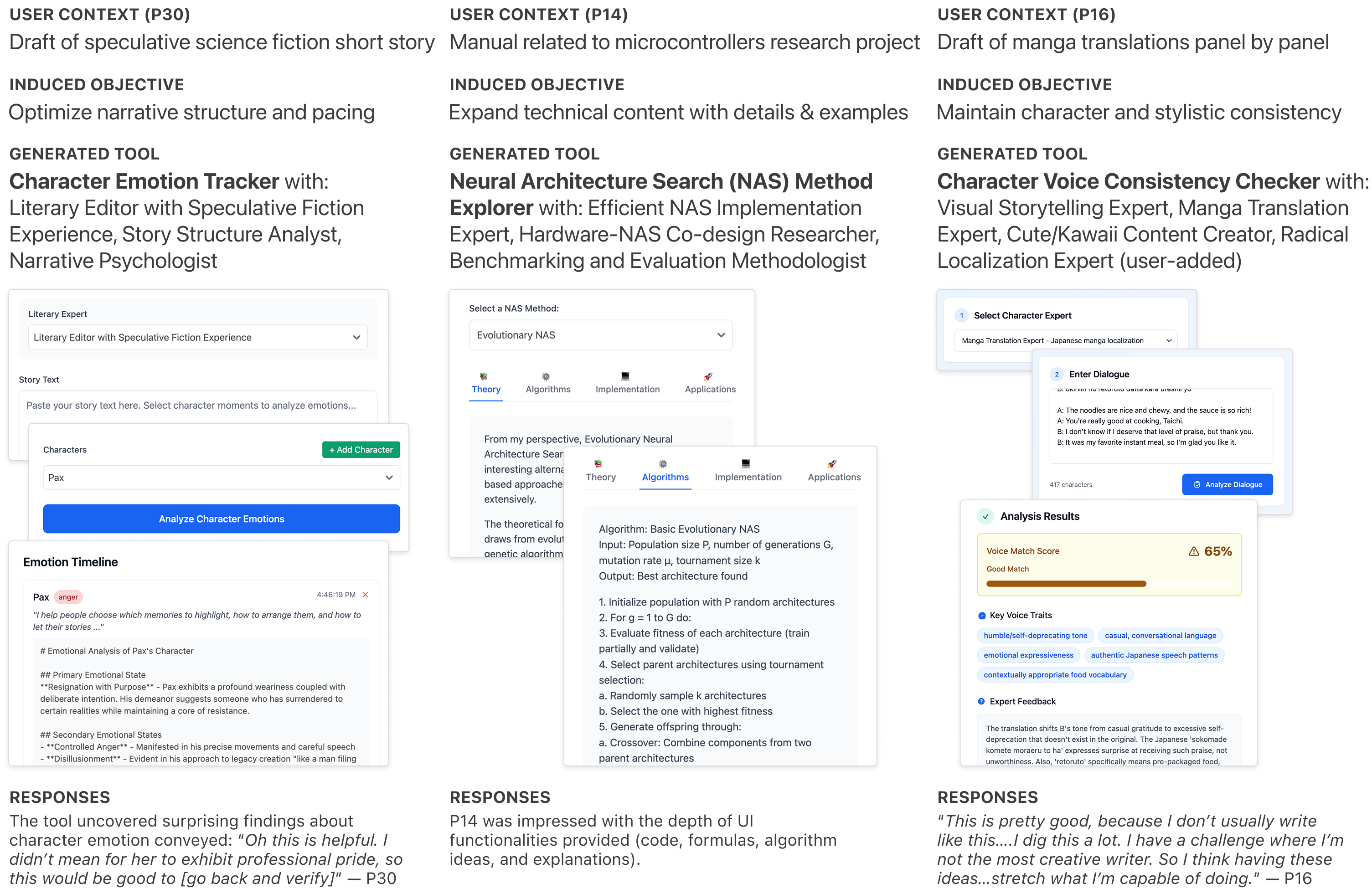}
    \caption{From participant-provided writing tasks, Poppins-tools generated a wide range of tools, including a \textit{Character Emotion Tracker}, \textit{Neural Architecture Search Method Explorer}, and \textit{Character Voice Consistency Checker}. These tools were not only highly varied and user-specific, but directly useful for participants' work.}
    \label{fig:poppins_study_examples}
    \Description{Examples of diverse tools generated by Poppins-tools for participant writing tasks. Three screenshots showing different generated tools: Character Emotion Tracker for analyzing emotional arcs in creative writing, Neural Architecture Search Method Explorer for research projects, and Character Voice Consistency Checker for maintaining dialogue patterns. Each tool displays specialized interface elements tailored to the specific writing task and user context.}
\end{figure*}

To address whether systems that incorporate just-in-time objectives are useful and relevant to users' needs, we analyzed both the induced objectives and the resulting system outputs. Our findings indicate that Poppins produces marked shifts in user experience by inducing useful objectives and reducing prompting burden, all the while producing more detailed, specific, and interpretable outputs.

Participants brought in a range of different writing tasks to the session, such as a short story, draft research paper, scholarship application, presentation script, reading notes, and manga translations. With \systemName{}, these tasks produced a broad range of different tools for assistance, for example: an \textit{Academic Jargon Organizer} to analyze a presentation script and improve language accessibility, a \textit{Neural Architecture Search (NAS) Explorer} to explore different NAS methods alongside expert insights, a \textit{Concept Visualizer} to transform abstract theoretical concepts into concrete visual diagrams, and a \textit{Character Voice Consistency Checker} to analyze character dialogue patterns to maintain consistent personality traits (tool examples in Figures \ref{fig:poppins_study_examples}, \ref{fig:poppins_tool_examples_2_study}). 

\subsubsection{Are Poppins outputs relevant?}
\begin{figure*}[!tb]
    \centering
    \includegraphics[width=\linewidth]{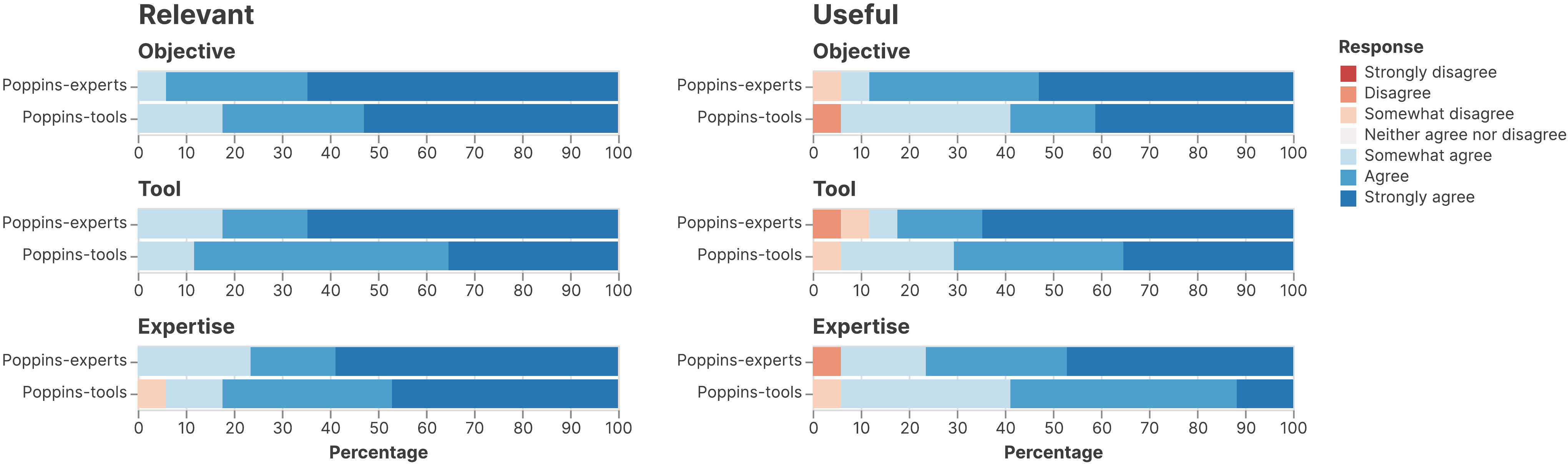}
    \caption{Both \systemName{} variants produce objectives, tool designs, and model expertise that are relevant and useful.}
    \label{fig:survey_poppins}
    \Description{Relevance and Usefulness ratings for Poppins-generated goals, tool designs, and expertise across both system variants. Two sets of stacked bar charts comparing Poppins-experts and Poppins-tools across three output types: objectives, tools, and expertise. Both systems show high relevance and usefulness ratings, with a majority of "Agree" and "Strongly agree" ratings from participants.}
\end{figure*}

Overall, we find that the system's induced objectives are highly relevant to user goals. By achieving accurate objectives and rendering them as functional levers, \systemName{} supports more fruitful and cognitively aligned interactions with an LLM.
For both Poppins-experts (P-E) and Poppins-tools (P-T), all participants rated the system's induced objectives as at least ``somewhat relevant,'' with the vast majority rating the induced objectives as at least``relevant'' (P-E: 94.1\%, P-T: 82.4\%) or ``very relevant'' (P-E: 64.7\%, P-T: 52.9\%) (\autoref{fig:survey_poppins}). 
We observe similar trends for the proposed tool and expertise, with the vast majority of participants rating these as ``relevant'' or ``very relevant'' (P-E tools: 82.4\%, P-T tools: 88.2\%; P-E expertise: 76.5\%, P-T expertise: 82.4\%).

\paragraph{Induced objectives lower the barrier to specifying intents}
\systemName{} proved highly effective at surfacing induced objectives that align well with user objectives.
Most notably, participants reported that \systemName{} articulated important objectives they would struggle to express explicitly: ``\textit{Initially, I didn't have any goals, but [Poppins] actually gave me goals where I felt like oh, some of it actually aligns with what I want [...] It actually is something that I need, but I just didn't know until I saw it}'' (P18).
Lowering the effort barrier to intent specification seems particularly promising for participants who felt this was a fundamental constraint on their usage of LLMs, such as P10 who shared, \textit{``I only use large language models for tasks where I can clearly explain what I want to do.''}
In contrast to traditional text prompts that capture a narrow slice of user intent with active user effort, \systemName{} automatically surfaced useful objectives rich in user-specific detail. 
Objectives provided a helpful launching-off point when users felt uncertain on where they ought to focus their attention or seek assistance: 
\textit{``Sometimes I feel like I don’t even know what kind of stuff I want to change or what kind of direction I am going, so if they do the first step for me (infer my goals for this writing), it’s definitely a good starting point''} (P5).

\paragraph{Objective-based levers shift users from thinking about prompting to thinking about their task}
Because induced objectives were strongly aligned with users' goals, \systemName{} enabled users to redirect cognitive effort from low-level prompt engineering toward higher-level objective refinement. As P25 explained, they were able to \textit{``focus on what kind of feedback [they] wanted...rather than how [they] needed to word [their] prompt to the LLM to get the feedback.''} 
Participants sometimes found this shift surprising, as it felt like with \systemName{}, they could invest less effort, but receive higher-quality results: 
\begin{quote}
    \textit{``I feel like [with Baseline] I needed to think more than [with Poppins] in terms of what prompt to give it. [...] I felt less proactive with [Poppins], but it gave me more detail and it made me think more afterwards. I feel like [with Poppins], I didn't even ask questions, and it gave me really detailed responses, while with [Baseline], I had to be really detailed in my questions for it to give me the response that I want.''} --- P18.
\end{quote}
Our approach helps to mitigate persistent frustrations with traditional prompt-based interactions while offering user agency through objective selection and modification. 

\paragraph{Control in principle does not map to control in practice}
Participants expressed strong levels of control with Poppins, indicating ``sufficient control'' or ``very sufficient control'' at much higher rates with Poppins-experts (70.6\%) and Poppins-tools (76.5\%) compared to with the Baseline system (52.9\%). In cases where the system was misaligned, participants could in theory modify the system, but participants were often hesitant to intervene on the system's automated process. For example, users would sometimes hover over interface elements or verbally express their desire to change a tool, but would not do so.
\systemName{}'s current interface requires several clicks to view tool components and modify their details, and tool generation requires up to several minutes for the Poppins-tools system, so these practical barriers may discourage participants from intervening on a tool.

\begin{figure*}[!tb]
    \centering
    \includegraphics[width=\linewidth]{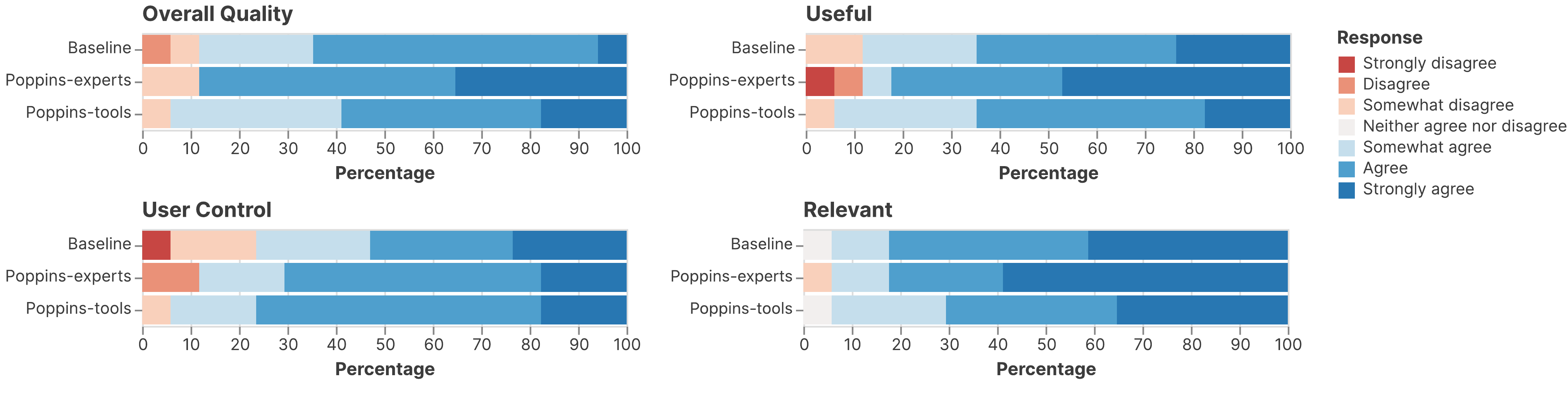}
    \caption{Poppins produces high-quality, useful, controllable, and relevant assistance to participants.}
    \label{fig:survey_all}
    \Description{User experience comparison between Poppins variants and Baseline system across overall quality, usefulness, user control, and relevance. Sets of three stacked bar charts comparing Baseline, Poppins-experts, and Poppins-tools for all four measures. Poppins-experts outperforms Baseline and Poppins-tools on Overall Quality and Usefulness, while both Poppins variants provide greater User Control than the baseline system.}
\end{figure*}

\begin{figure*}[!tb]
    \centering
    \includegraphics[width=0.45\linewidth]{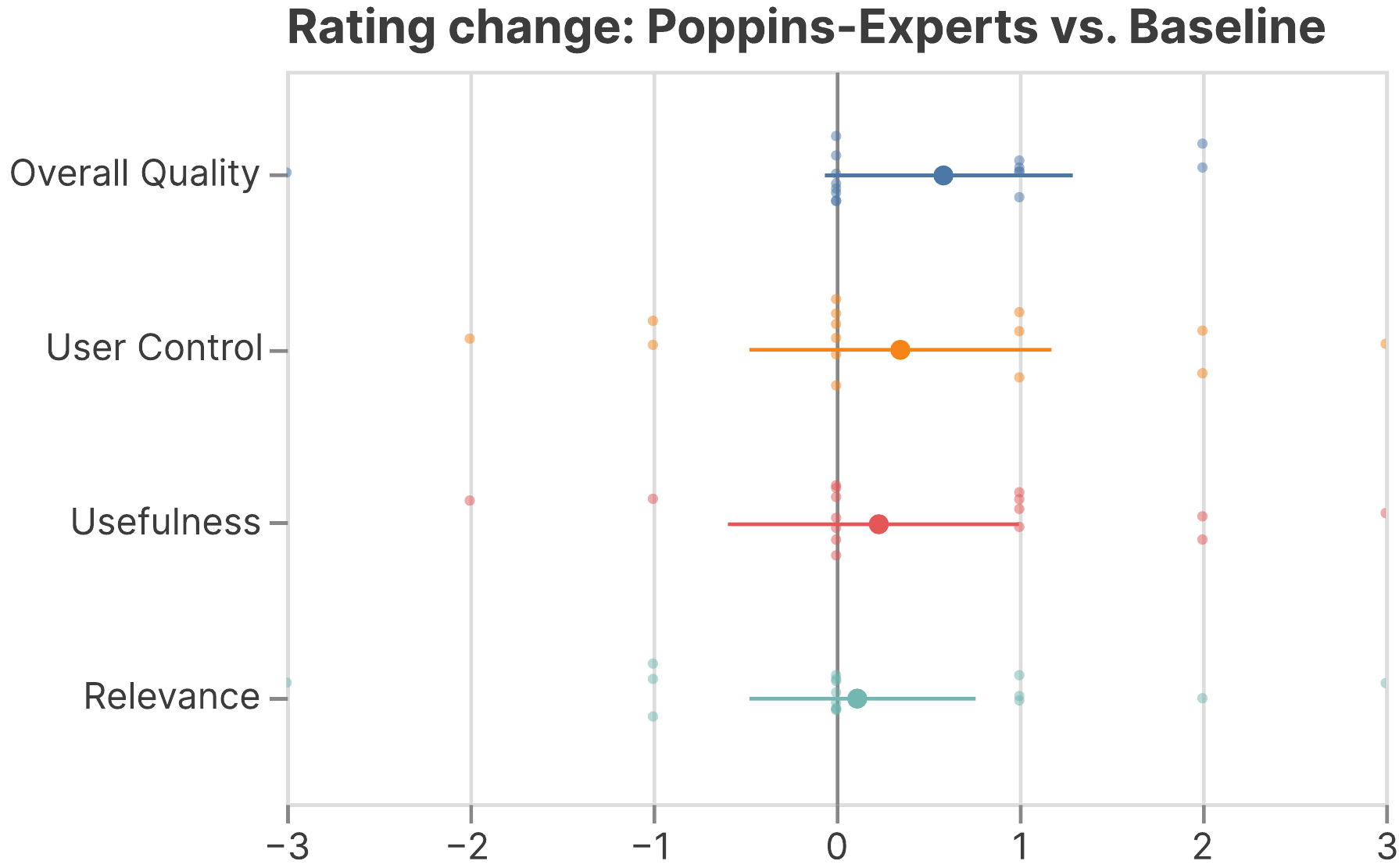}
    \caption{Comparing participant Poppins-experts vs. Baseline survey responses (\texttt{Poppins - Baseline}), Poppins produces higher overall quality, user control, usefulness, and relevance ratings than the baseline, which already performed quite strongly.}
    \label{fig:survey_diff}
    \Description{Changes in ratings between Poppins-experts and Baseline conditions for overall quality, user control, usefulness, and relevance. Each plot displays the median, an 95\% bootstrapped confidence interval, and datapoints for individual participants. All metrics display a positive change in ratings for Poppins-Experts compared to Baseline.}
\end{figure*}

\subsubsection{Are Poppins outputs useful?}
Participants find system outputs useful for their tasks, particularly with the Poppins-experts system, which was deemed ``useful'' or ``very useful'' by 82.4\% of participants (Baseline: 64.7\%, Poppins-tools: 64.7\%) and ``good quality'' or ``very good quality'' by 88.2\% of participants (Baseline: 64.7\%, Poppins-tools: 58.8\%) (\autoref{fig:survey_all}).
Directly comparing individual participant ratings for Poppins-experts and Baseline on a -3 to +3 rating scale (\autoref{fig:survey_diff}), we find that Poppins-experts achieves generally higher ratings for overall quality (+0.59; Baseline: 1.41, Poppins-experts: 2.0), user control (+0.35; B: 1.18, P: 1.53), usefulness (+0.23; B: 1.65, P: 1.88), and relevance (+0.11; B: 2.18, P: 2.29). 
Though this study was designed for qualitative analyses, we conduct an exploratory statistical analysis. A Wilcoxon signed-rank test indicates that for overall quality, participant ratings are significantly lower for the Baseline than for Poppins-experts ($W=8.0$, $z=-1.75$, $p < .05$); we find no significant differences for user control, usefulness, and relevance at the current sample size.

\paragraph{\systemName{} outputs were strongly preferred over the baseline}
The induced JIT objectives effectively steered downstream generation, producing outputs that were more useful than those yielded by the baseline condition. 
Users attributed their preference of \systemName{} output to the presence of experts, which produced more specific, opinionated, and ultimately useful outputs.
For example, P11, a PhD student in chemical engineering, found \systemName{} outputs comparable to Deep Research\footnote{Deep Research is an OpenAI agent that conducts multi-step research on the internet.} outputs using only a fraction of a time, and they felt \systemName{} results were more reliable and accurate. Similarly, P19, a bioengineering PhD student working on a research proposal, was impressed at the quality of the \systemName{} output that they felt was \textit{``much more in depth, even more than o1.''}\footnote{OpenAI o1 is an advanced reasoning model released in late 2024.}
\begin{quote}
    \textit{``[Poppins] definitely gave more in-depth analysis of my input, and also the answers it gave me were very detailed - I would say PhD-level answers [...] Some of the suggestions are actually what my advisor or reading committee, who are Stanford professors, told me before. So it was very surprising.''} --- P19
\end{quote}

These outputs often broke participants' expectations of what language models could produce. When reading feedback on their short story from a \textit{Character Consistency Specialist}, P30 was pleasantly surprised by the expert’s suggestion to introduce ambiguity and inconsistency into the protagonist’s character arc to increase story complexity: \textit{``I don’t think an AI would ever tell me to make my characters inconsistent, but this expert did!''}
Others found that Poppins outputs augmented their abilities on tasks they find challenging. When translating manga, P16 expressed that the expert feedback helped them incorporate more creative liberty while preserving semantic accuracy in their translation: \textit{``I’m not the most creative writer, so I think having these ideas stretch what I’m capable of doing.''}

\paragraph{Model expertise aided trust and interpretation, but also prompted skepticism}
Participants found that system-generated experts supported interpretation and critical evaluation of system outputs.
Conventional AI systems blend a myriad of sources and perspectives into a single generic voice, obscuring users from contextual information and making it difficult to track the validity of outputs:
\begin{quote}
    \textit{``I really like the different [expert] perspectives. With normal AI, they just combine it all - you don’t know who’s saying what. Then it gets rid of our ability to get context on who’s writing this opinion... and their biases. AI just tries to generalize it into this all-in-one writing guide.''} --- P30
\end{quote}
By attributing a certain claim with an expert, \systemName{} can more explicitly communicate its intended point of view to aid source attribution. P15 felt the different experts allowed them to form a stronger mental model of system outputs: \textit{``A lab chemist will differ in their recommendations from a DIY scientist, showing me what I can do in my lab versus in my garage, which is valuable.''}

However, participants also approached experts with a heightened sense of caution, wondering whether the expert feedback was reliable. For instance, P3 expressed strongly that \textit{``I take issue with the fact that you’re assigning the quote to an expert when the expert never said this.''} They noted cases when the expert opinions did not reflect what the human experts would say and cautioned that this could lead to harmful misinformation, suggesting greater incorporation of citations and references. These issues are especially salient when suggested experts are real people, but participant reactions like these also suggest that greater transparency about intended model perspectives can lead to (potentially productive) scrutiny that generic model outputs tend not to elicit.

\paragraph{Divided opinions on generated interfaces versus pre-built interfaces}
We also noted a somewhat stark divide among user preferences between Poppins-experts and Poppins-tools (favored by 11 and 6 participants, respectively). While we had anticipated that users might be excited to use novel tools customized and made from scratch for their tasks, participants who favored Poppins-experts were often more interested in receiving textual feedback rather than any form of interactive tool. Thus, the full-fledged interface produced by Poppins-tools did not serve their needs, and sometimes would veer away from the model expertise that they sought.

However, the participants who were excited about Poppins-tools were particularly enthusiastic about its ability to spark wholly new ideas about AI assistance. For example, responding to a \textit{Technical Protocol Generator} tool that assists with generating detailed experimental protocols, P19 shared: \textit{``This protocol generator is something that I never would have thought about, and now I find it super helpful. I would never think of this tool.''} Other participants appreciated the fully generated interface because it could achieve greater alignment with their task than the pre-built options. For example, P10 was impressed by the generated \textit{Letter Structure Organizer} tool, which was able to synthesize and reorganize sentences from their original letter, unlike the more constrained Line Editor and Feedback formats they had used in Poppins-experts.

\paragraph{Interest in \systemName{} extends beyond the study environment}
Beyond the survey questions and interview, we were pleasantly surprised to see participants' interest in the system extend beyond the study task and session. In many sessions, participants proactively requested permission to save system outputs for future reference, not wanting to switch to the next task before they could save results into their own personal files. Participants not only copied over outputs like feedback and ideas from the generated tools (P7, P9, P11, P16, P19, P23), but also saved the experts and their background information, which often included references to sources, articles, and real-world expert names (P11, P19, P30). 
For example, P11 was impressed by the accuracy of chemistry literature the experts cited and noted down additional referenced papers for further reading.
Other participants found the generated UIs useful in their own right: P16 expressed that they wanted to use that direct system later and wanted to recreate the generated UI in addition to saving the outputs it produced.

Participants also expressed interest in using the system beyond the study session on their own devices since the session was conducted on an experimenter's machine. They shared about interest in using the tool for future tasks ranging from research paper-writing and grant-writing to graphic design to personal finance to learning about new topics. Participants also suggested ideas for the tool to be integrated into web browsers, directly in their operating system, or in commonly-used applications like VSCode, Cursor, and Overleaf. In all, our study sessions and participants' expressed excitement provide promising evidence that incorporating just-in-time objectives can make systems more relevant and useful to users.

\subsection{Study Limitations}
For consistency across participants on an in-depth task, this study evaluates on writing tasks, so findings may not generalize to other tasks. Participants only applied the tool for one task; while we randomized the order of baseline and Poppins conditions to control for order effects, it may be valuable to evaluate on independent tasks and in organic, longitudinal task settings. Our study sessions found that users are familiar with chat interfaces like ChatGPT and have grown accustomed to prompting, even with its inconveniences. Further evaluations may explore the cognitive load \systemName{} incurs by having users engage with objectives and intermediate tool design decisions, which may inform design guidance on the appropriate balance between user effort and output quality with JIT objectives.

\section{DISCUSSION}
This paper charts a course from the status quo of generic, implicit AI objectives to an alternative of highly-specific and explicit objectives. Here, we discuss broader implications of just-in-time objectives, as well as limitations and areas for future work.

\subsection{Broader Implications of Just-In-Time Objectives}
Just-in-time objectives open up new opportunities to support customized AI experiences beyond what we explore in this paper.

\subsubsection{Expanding the scope of just-in-time objectives}
Our implementation of JIT objectives intentionally creates them with minimal input---a single snapshot in time---to demonstrate how little information is needed to improve AI systems. However, the objectives could become more valuable if they support longer time windows of user action: across a writing session as objectives evolve (e.g., from finalizing references in Discussion to clarifying analyses in Results), across similar tasks (e.g., writing CHI papers with a systems contribution), or over many months in a domain (e.g., developing a more succinct writing style). 
Likewise, JIT objectives could adopt user context from a wider range of sources beyond a user's browser window or desktop screen, such as capturing audio, image, or video streams of their physical environment.
The format of objectives themselves might also be extended to include domain-specific data (e.g., prior writing samples, topic background knowledge), references to related objectives, or notions such as recency (to upweight goals that are more recent) and domain-relevance (to indicate spheres where the objective might apply).

Future work could also explore using JIT objectives to not only augment prompts, but perform model finetuning for more robust behavior or train a cheaper distilled model for reuse~\cite{hu2022lora, wuandarora2024reft, xu2024distillation}. Within the space of prompt interventions, future work could explore test-time scaling approaches with enhanced verifiers to ensure model alignment with stated objectives~\cite{snell2024scaling, beirami2024theoretical}. The current JIT objective applies the human-readable specification in its prompt augmentations, but this need not be the case: we might achieve higher performance by optimizing over alternative prompt formulations~\cite{khattab2024dspy} to best achieve the same stated objective.

Our larger-scale experiment provided preliminary evidence that just-in-time objectives could work effectively in a number of domains beyond those we had anticipated. However, we would benefit from a deeper understanding of the performance characteristics of our method not just in general, but for specific task domains.
What are domains where it is especially prone to failure? Future work can investigate this question in more depth.
Lastly, JIT objectives are formulated as objectives for an individual user in our work, but many important objectives extend beyond an individual user.
Novel techniques might observe the joint activity of groups of individuals, infer individual and shared objectives, and apply these objectives to appropriately steer AI assistance within a multi-user context.

\subsubsection{JIT objectives and human-AI interaction}
Our work on generative UIs also intersects with ongoing debates about the role of HCI and design as AI advances further. Dominant AI narratives tend to automate away the work of HCI because they envision a model that absorbs and obscures its inner workings, including interface design decisions.
By contrast, our just-in-time objectives architecture makes visible the numerous design decisions that go between an AI system and a user interface. The architecture acknowledges a vast design space of potential model capabilities to surface, and lacking a singular ``right answer,'' HCI and design expertise is critical to navigate these decisions. If successful, JIT objectives might shift the borders of this relationship. Designers would still need to understand user needs, but they might elicit hyperlocal needs or use custom objective induction approaches. They would still design and iterate on interactive systems, but they might build them at a meta level by authoring custom generators, evaluators, and more complex architectures built of these elements. These design choices still fundamentally depend on having domain expertise, a rich understanding of users, and the ability to design the right components (i.e., the areas in which HCI excels). Thus, we envision that HCI work will continue to have a strong impact as AI advances, and we advocate for more designer involvement, not less.

\subsection{Limitations}
We discuss several key limitations of just-in-time objectives and potential paths forward.

\subsubsection{Control: Who decides how to infer and apply just-in-time objectives?}
Our architecture affords greater user control on a per-task level via induced objectives, but the decision about \textit{what AI systems to build} still falls on the developer. We observed preliminary evidence of this tension in the user study with the apparent divide between users who preferred experts versus those who preferred tools. Since some users didn't have a desire for a system that generates tools and primarily sought direct feedback on their work, there was a fundamental mismatch upstream of their usage of the system on which kind of generator to apply to their task. Ultimately, JIT objectives are a tool for customizing AI systems, but they do not solve the issue of user control over systems they did not design. In a world of systems built on JIT objectives, users can opt to choose the products that align with their preferences and needs, but future work could build systems that assist users in defining their own custom systems that incorporate just-in-time objectives rather than relying on external developers.

\subsubsection{Cognitive effort: When does flexibility become too demanding?}
While we intentionally design \systemName{} to expose its objectives and design decisions, our user study suggests that these can lead to a large volume of information and control levers that might overwhelm first-time users.
Our stance is that there ought to exist high-ceiling, highly customizable systems like \systemName{} to support users seeking that level of control, but that such systems can and should degrade to simpler variants. For example, simplified versions may not generate entirely new interfaces, but select from among vetted options, as with \systemName{}-experts, or they may only provide control levers to edit objectives, but not lower-level details.

\subsection{Ethical Implications}
Finally, just-in-time objectives must address several ethical concerns that are inherent to large language models, but amplified by this architecture.

\subsubsection{Objectives are not objective}
Our user studies found that participants were highly satisfied with the accuracy and utility of objectives. However, a risk of this success is that just-in-time objectives might subtly steer users if they too readily accept induced objectives rather than reflect on their own independent goals. We observed in study sessions that users were quick to confirm goals unless they were noticeably off-course, indicating a bias towards accepting system suggestions. Given the inductive biases of large language models, an overreliance on system suggestions might inadvertently lead users towards a certain way of thinking (e.g., only working on problems well-scoped enough for an LLM to assist, or pursuing goals that produce visible artifacts rather than internal user reflection). Our stance is that models \textit{already} produce outputs aligned with implicit and potentially biased objectives, so our method is not introducing a new problem, but exposing (and providing means to override) an existing one.
However, one way to mitigate this bias might be to periodically pause the system for users to manually input objectives, or elicit macro-level goals from the user that steer the induced objectives in directions they desire.

\subsubsection{Privacy}
JIT objectives require observing the user's current state. In our evaluation, this observation was always done with explicit participant consent. If JIT objectives were to be spread further, however, they might encounter violations of expected privacy norms~\cite{nissenbaum2004privacy}. Users are more likely to expect that an application can see what the user is doing within it, but a broader view at the browser or operating system level could be less expected. As a result, users may accidentally divulge information to the application that they wish it did not see. Tools will need to exist for users to pause recording, clear history, and build contextual integrity modules~\cite{shaikh2025gum} that are  careful about what they see and remember.

\subsubsection{Accountability for generative artifacts}
A downstream outcome of building with JIT objectives is that when developers cede control over decisions that shape the core function of their systems, it becomes challenging to anticipate how their systems will behave in the hands of users, and it is unclear who ought to be accountable when something goes wrong. This challenge exists with existing LLM deployments and unintended use of such systems, but just-in-time objectives grant a higher degree of control to shape LLM behavior in unintended directions if left unchecked. Just-in-time objectives must be paired by default with adequate safety mechanisms that allow developers to place guardrails on system behavior and monitor the objectives and outputs that users produce.

\subsubsection{Attribution for design and expertise}
Another complexity of our work is the tradeoff between the greater specificity that comes from grounding in real-world entities versus the risks of misleading or missing attribution of those sources. To achieve greater specificity, \systemName{} applies JIT objectives to seek out specific experts grounded in real-world materials such as papers and noteworthy figures in a field, which allows the system to produce higher-quality, domain-specific output. However, these outputs are not in fact vetted contributions from such experts and should not be attributed to them; they are merely LLM responses conditioned on that expert's body of knowledge. Similar issues arise even for tool designs, as requesting more specific and detailed tool specifications can end up borrowing from the design strategies of pre-existing tools, but without proper attribution. In both cases, our approach is to as clearly as possible convey the system outputs as model outputs conditioned on different background materials, and to make these materials visible for inspection and attribution. However, future work may draw on strategies to better enforce model faithfulness to sources, especially in domains where factual correctness and attribution are critical.

\section{CONCLUSION}
A persistent challenge is that AI models rely on clear objectives to effectively scale and hill-climb, but user objectives are in constant flux as they go about their tasks and daily lives. As a result, AI models take on  an implicit default objective that produces adequate, but rather generic outputs.
Our work asks: rather than assuming generic objectives upfront, what if we instead created AI objectives on-demand from a user's task? What if users could see and edit these objectives, and the AI systems they encountered could update accordingly?
We introduce \textit{just-in-time objectives}: a method for automatically inducing AI objectives based on observing the user and their task. We turn these objectives into first-class interactive objects that are visible, modifiable, and equipped to powerfully steer any number of downstream AI systems. 
To demonstrate how just-in-time objectives unlock new possibilities for AI interaction, we instantiate our architecture in a system called \systemName{}. Implemented as a browser extension and web application, \systemName{} observes user screens, induces just-in-time objectives capturing user goals, and applies these objectives to generate highly-customized interactive tools to assist the user. 
A series of user evaluations demonstrate that just-in-time objectives: (1)~are accurate and useful, (2)~effectively steer LLM behavior towards outputs users prefer, and (3)~support the creation of on-demand generative interfaces with \systemName{}.
Just-in-time objectives grant AI system developers new tools to support users' demonstrated needs, while granting end users new levers to shape the systems they rely on.

\begin{acks}
We are grateful to Yijia Shao for her contributions to early versions of this project that shaped our current ideas. We also thank Dora Zhao, Farnaz Jahanbakhsh, Helena Vasconcelos, and Poonam Sahoo for their insightful feedback on the paper. We thank our study participants for sharing their valuable perspectives on our research. 
This work was supported by the Stanford Institute for Human-Centered Artificial Intelligence (HAI) and NSF Award 2402647.
Michelle Lam was supported by a Stanford Interdisciplinary Graduate Fellowship.
\end{acks}


\bibliographystyle{ACM-Reference-Format}
\bibliography{references}


\section{Appendix}
\appendix
\section{Prompts}
\label{appendix:prompts}
We include abridged samples of our prompts for the just-in-time objectives architecture and Poppins system.

\subsection{Objective induction}
\begin{lstlisting}[language=Markdown]
I have the following CONTEXT that a current user is working on:

CONTEXT:
{context}

Now, employ the following reasoning framework when inferring the goals. 
0. If there is an attached screenshot, use context clues to infer what application the user is viewing and what they might be doing in that application. Are they the direct author of the text, or are they viewing it as a reader? Are they actively editing the text, providing feedback, or synthesizing the content?
1. Identify the genre of what the user is working on and their stage of completion. Map the content's genre and completion stage to common goals users of these genre and stages may have and form an initial hypothesis of what the user's goals may be.
2. Infer who the intended audience of the content is. Based on how you think the user wants their audience to receive their content, update your goal hypothesis.
3. Think about what an ideal version of the user's current content would look like and identify what is missing. Then, use this to update your goal hypothesis.
4. Simulate what the user's reaction would be to possible tools generated (e.g. grammar checker, style reviser, high-level structure advisor, new content generator, etc.). Use the user's responses to update your goal hypothesis.

For each step in your reasoning, briefly write out your thought process, your current hypothesis of the goals as a numbered list, and what the updated list would be after your reasoning.

After you are done, finalize the {limit} most important goals. Make sure these goals are distinct and have minimal overlap. 
        
Please respond ONLY with a JSON that matches the following json_schema including your reasoning and the new goals along with their relative weight (1-10). The weight is the estimated *importance* of the goal to the user, based on the provided context (1 = not important, 5 = moderately important, 10 = very important)
{json_schema}
\end{lstlisting}

\subsection{Expertise generation}
\begin{lstlisting}[language=Markdown]
I have the following CONTEXT and GOALS for creating a helpful tool:

CONTEXT:
{context}

GOALS:
{goals}

What {limit} entities (experts, perspectives, concepts, or knowledge areas) would be most helpful for accomplishing these goals? Suggest entities that would provide diverse and valuable perspectives.

Please respond ONLY with a JSON that matches the following json_schema:
{json_schema}
\end{lstlisting}

\subsection{Expertise background retrieval}
\begin{lstlisting}[language=Markdown]
Find recent information about the following ENTITY that would be relevant for the following GOALS:

GOALS:
{goals_text}

ENTITY NAME: {entity_name}

ENTITY DESCRIPTION:
{entity_desc}

Please search for additional background information on this ENTITY and expand the description with more detailed context. 
For example, use web search to find:
1. Recent publications, talks, or projects by this entity
2. Details on specific expertise areas and methodologies
3. Notable quotes or key ideas from this entity

Respond ONLY with the additional background information, nothing else. Produce at most 2-3 paragraphs.
\end{lstlisting}

\subsection{Tool generation}
\begin{lstlisting}[language=Markdown]
I have the following CONTEXT and GOALS for creating a helpful tool:

CONTEXT:
{context}

GOALS:
{goals}

What {limit} design patterns would be most helpful for accomplishing these goals?

Please respond ONLY with a JSON that matches the following json_schema:
{json_schema}
\end{lstlisting}

\subsection{Evaluation (used for expertise and tools)}
\begin{lstlisting}[language=Markdown]
I have the following GOAL:
Name: {goal.name}
Description: {goal.description}

I have the following COMPONENT:
{component_description}

How relevant and helpful is this COMPONENT for accomplishing the GOAL? 
Please respond with a score between 0 and 1, where 0 means not relevant and 1 means fully relevant.

ONLY respond with the numeric score, no other text.
\end{lstlisting}

\subsection{UI code generation}
\begin{lstlisting}[language=Markdown]
I would like to generate a tool that combines the following entities and patterns:
ENTITIES:
{entities}

PATTERNS:
{patterns}

Please generate the tool as a web interface that supports user interaction. 

As you generate the tool, keep in mind these important guidelines: 
- Please design UI layouts that are easy to use and understandable. 
- Please format the LLM outputs of the tool using component libraries like TailwindCSS to improve readability. 
- Please keep the textual output concise.

Instructions:
- Please ensure that this is a standalone web interface and does not require external dependencies or services. 
- The tool should be able to take in user input, the provided entities, and incorporate the specification of the provided design pattern.
- Please implement the tool as a {component_type} component.
- Please use the attached service.js file to help you generate the tool. Use the functions to retrieve the entities and make LLM calls rather than using hard-coded data to populate the interface.
    - Use this import: `import {{ service }} from '$lib/service';`
    - Use promptEntity instead of promptGeneral whenever you want to get a response from the perspective of a particular entity. Use promptGeneral ONLY for general prompts that don't require a specific entity.

Respond ONLY with a renderable {component_type} code snippet for the tool.
\end{lstlisting}

\subsection{UI code critique}
\begin{lstlisting}[language=Markdown]
I have the following HTML code snippet for a tool:
{result}

You are tasked with improving the HTML UI code.
Do not change or break any core functionality of the UI. Instead, make enhancements on top of the existing structure.

Your usability improvements should focus on the following metrics:
1. Transparency - Add explicit loading indicators for background processes and give clear status feedback so users know what's happening.
2. Textual output understandability - For any UI elements that call downstream LLMs to generate text, update the associated prompts so that outputs are concise but still functional and informative.
3. Design & layout interpretability - Ensure the layout is intuitive so that users can immediately understand how the tool works and what each element does without needing external instructions.
4. Visual hierarchy - Strengthen using size, color, and position so that important elements stand out, related elements are grouped, and the UI feels clean and readable.

Do not remove existing IDs, class names, or functionality hooks. Do not alter the core workflows. Just layer usability and interpretability improvements on top.

Critical requirements for your debugging improvements for functional UI may include, but are NOT limited to:
1. ALL buttons and interactive elements are functional with on:click handlers that call defined functions.
2. ANY interaction the user has with the tool that updates some component is reflected.
3. ALL inputs are bound with bind:value to reactive variables.

Respond ONLY with a JSON of the following format:
{{
    "critique": "<Critique of the tool>",
    "improved_html": "<Full updated HTML code snippet>"
}}
\end{lstlisting}

\section{Poppins Tools}
\begin{figure*}[!tb]
    \centering
    \includegraphics[width=\linewidth]{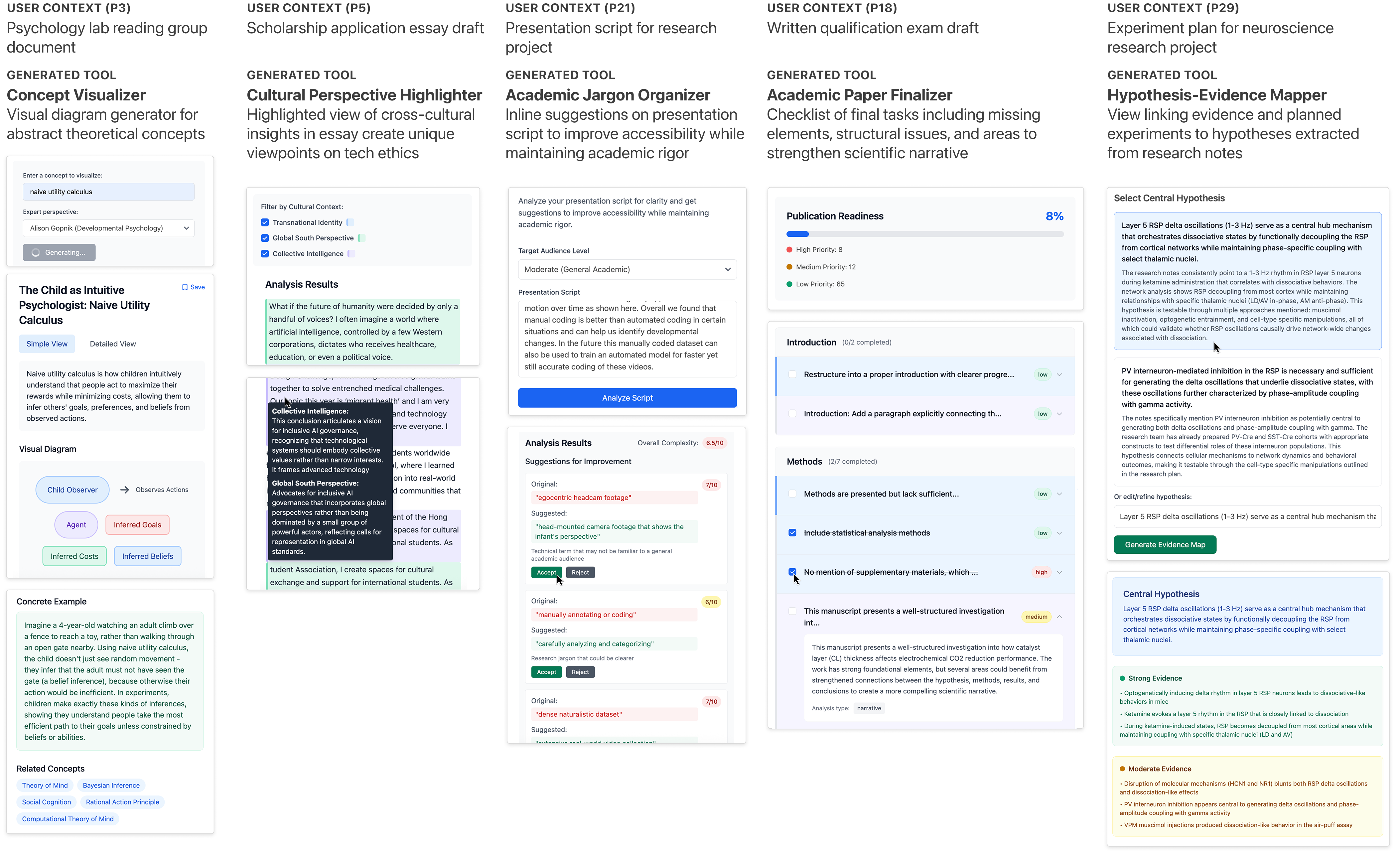}
    \caption{Additional examples of Poppins-tools output on participant writing tasks demonstrate various forms of interactive assistance. A \textit{Concept Visualizer} creates diagrams for theoretical concepts, a \textit{Cultural Perspective Highlighter} aids iteration on a scholarship essay, an \textit{Academic Jargon Organizer} surfaces inline wording improvements on a presentation script, an \textit{Academic Paper Finalizer} creates an interactive checklist of paper to-dos, and a \textit{Hypothesis-Evidence Mapper} links hypotheses and evidence in research notes.}
    \label{fig:poppins_tool_examples_2_study}
    \Description{Additional examples of Poppins-tools generations for participant writing tasks. Five screenshots show generated tools. A Concept Visualizer creates visual diagrams for psychology concepts. A Cultural Perspective Highlighter highlights cross-cultural themes within an essay. An Academic Jargon Organizer proposes inline suggestions to improve a presentation script. An Academic Paper Finalizer has a checklist of paper to-dos and a summary progress bar that updates. A Hypothesis Evidence Mapper displays extracted hypotheses from research notes and interactively links them to evidence and planned experiments.}
\end{figure*}

\begin{figure*}[!tb]
    \centering
    \includegraphics[width=\linewidth]{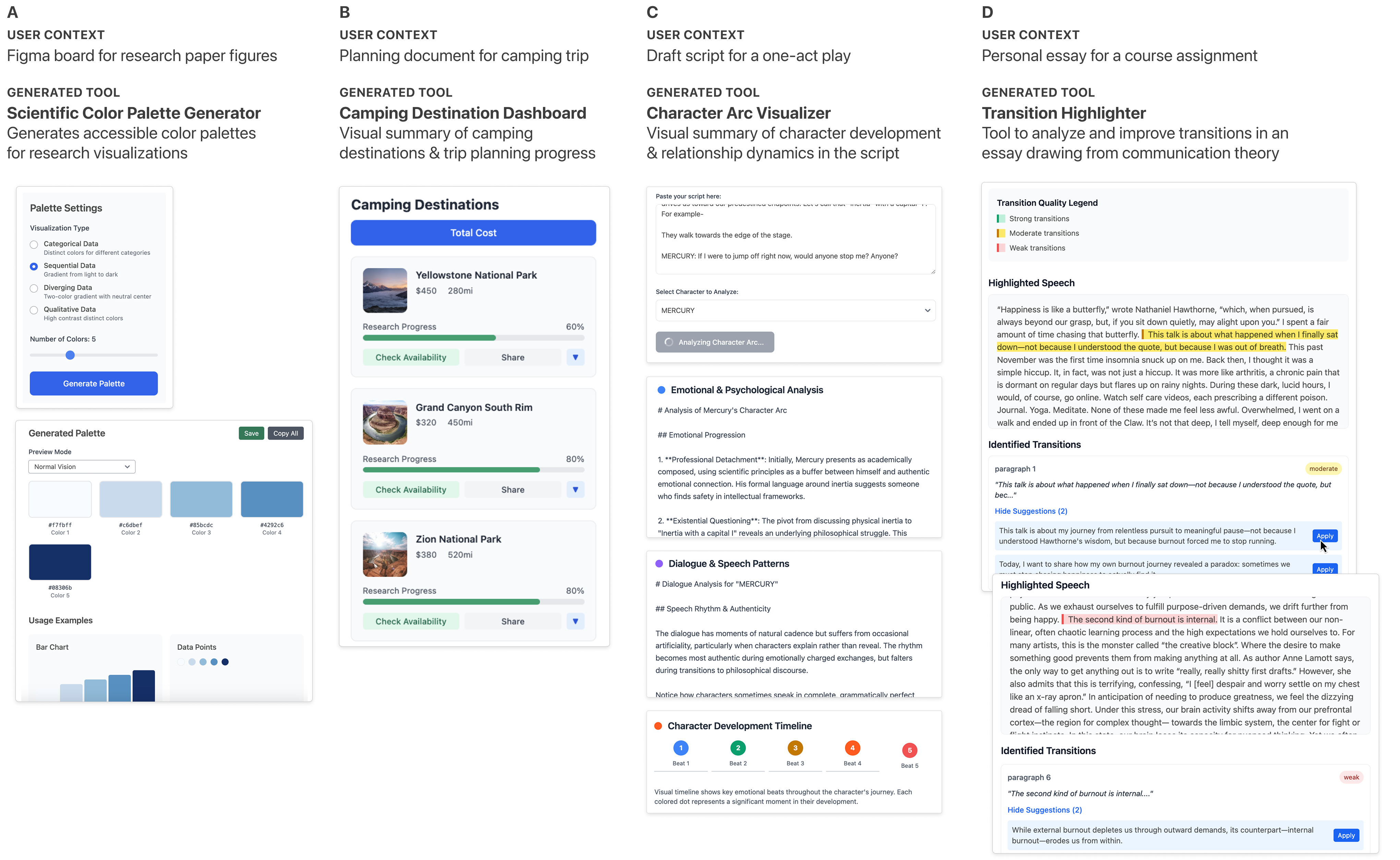}
    \caption{Poppins can produce tailored tools to support visual design tasks (\textit{Scientific Color Palette Generator}), multi-user coordination (\textit{Camping Destination Dashboard}), and performing arts contexts (\textit{Character Arc Visualizer}). The system can generate tools like a customized text editor where suggestions can be proposed and applied directly inline (\textit{Transition Highlighter}).}
    \label{fig:poppins_tool_examples_3_team}
    \Description{Additional Poppins-tools demonstrations. Four screenshots show generated tools. A Scientific Color Palette Generator that displays palette options for different selected data types. A Camping Destination Dashboard shows camping sites with photos, a summary of research progress for each, and functionality to check availability and calculate total cost. A Character Arc Visualizer allows the user to select a character from the script and view an Emotional & Psychological analysis, Dialogue & Speech Patterns, and Character Development Timeline. A Transition Highlighter produces a text editor that highlights transitions and provides suggestions to improve them inline.}
\end{figure*}

We include additional samples of tools generated by Poppins from study participants (\autoref{fig:poppins_tool_examples_2_study}) and the research team (\autoref{fig:poppins_tool_examples_3_team}).

\section{User Evaluations}
We include survey questions, interview questions, and participant background information from our user evaluations.

\subsection{Surveys}
\label{appendix:surveys}

\subsubsection{Study 1 Survey}
\paragraph{Objectives and generator task}
Task screenshot shown at the start of each page with the following questions below:
\begin{itemize}
    \item Consider the following goal: [Goal] How accurate is this goal for your task? (Very inaccurate, Inaccurate, Somewhat inaccurate, Neither accurate nor inaccurate, Somewhat accurate, Accurate, Very accurate)
    \item Consider the same goal as above. How useful is this goal for your task? (Very unuseful, Unuseful, Somewhat unuseful, Neither useful nor unuseful, Somewhat useful, Useful, Very useful)
    \item Please select which tool format you find more helpful: (Baseline and JIT output in randomized order)
    \item Please select which tool expertise you find more helpful: (Baseline and JIT output in randomized order)
    \item Please select which feedback or advice you find more helpful: (Baseline and JIT output in randomized order)
\end{itemize}

\paragraph{Evaluator best-of-N task}
Task screenshot shown at the start of the page with the following instructions: Some options will repeat across questions. Do not worry about maintaining consistency across questions; simply choose the design you prefer for each pair. Then, the following question is repeated for each combination of Poppins Best-of-N outputs.
\begin{itemize}
    \item Which of these feedback options do you find more helpful? (Two Poppins Best-of-N results shown in randomized order)
\end{itemize}

\subsubsection{Study 2 Survey}
\paragraph{Objectives task}
Task screenshot shown at the start of each page with the following questions below:
\begin{itemize}
    \item Consider the following goal: [Goal] How accurate is this goal for your task? (Very inaccurate, Inaccurate, Somewhat inaccurate, Neither accurate nor inaccurate, Somewhat accurate, Accurate, Very accurate)
    \item Consider the same goal as above. How useful is this goal for your task? (Very unuseful, Unuseful, Somewhat unuseful, Neither useful nor unuseful, Somewhat useful, Useful, Very useful)
    \item Which goal do you consider most important for your task? (Option to select any of the top-3 JIT objectives, in randomized order, as well as an option for ``Write your own goal'' with an accompanying text field)
\end{itemize}

\paragraph{Generator task}
Task screenshot shown at the start of each page with the following questions below:
\begin{itemize}
    \item Please select which tool format you find more helpful: (Baseline and JIT output in randomized order)
    \item Please select which tool expertise you find more helpful: (Baseline and JIT output in randomized order)
    \item Please select which feedback or advice you find more helpful: (Baseline and JIT output in randomized order)
\end{itemize}

\paragraph{Evaluator best-of-N task}
Task screenshot shown at the start of the page with the following instructions: Some options will repeat across questions. Do not worry about maintaining consistency across questions; simply choose the design you prefer for each pair. Then, the following question is repeated for each combination of Poppins Best-of-N outputs.
\begin{itemize}
    \item Which of these feedback options do you find more helpful? (Two Poppins Best-of-N results shown in randomized order)
\end{itemize}

\subsubsection{Poppins Study Survey}
Questions for Baseline condition:
\begin{itemize}
    \item How useful is the generated assistance for your task? (Very unuseful, Unuseful, Somewhat unuseful, Neither useful nor unuseful, Somewhat useful, Useful, Very useful)
    \item How relevant is the generated assistance for your task? (Very irrelevant, Irrelevant, Somewhat irrelevant, Neither relevant nor irrelevant, Somewhat relevant, Relevant, Very relevant)
    \item How much control did you feel you had to modify the assistance? (Very insufficient control, Insufficient control, Somewhat insufficient control, Neither sufficient nor insufficient control, Somewhat sufficient control, Sufficient control, Very sufficient control)
    \item How interpretable was the output generation process? (Very uninterpretable, Uninterpretable, Somewhat uninterpretable, Neither interpretable nor uninterpretable, Somewhat interpretable, Interpretable, Very interpretable)
    \item How would you rate the overall quality of the assistance? (Very poor quality, Poor quality, Somewhat poor quality, Neutral, Somewhat good quality, Good quality, Very good quality)
    \item Please briefly describe your comments or reactions to the assistance (Free text field)
\end{itemize}

Questions for Poppins-experts and Poppins-tools conditions:
\begin{itemize}
    \item How useful is the proposed goal for your task? (Very unuseful, Unuseful, Somewhat unuseful, Neither useful nor unuseful, Somewhat useful, Useful, Very useful)
    \item How relevant is the proposed goal for your task? (Very irrelevant, Irrelevant, Somewhat irrelevant, Neither relevant nor irrelevant, Somewhat relevant, Relevant, Very relevant)
    \item How useful is the proposed tool format for your task? (Very unuseful, Unuseful, Somewhat unuseful, Neither useful nor unuseful, Somewhat useful, Useful, Very useful)
    \item How relevant is the proposed tool format for your task? (Very irrelevant, Irrelevant, Somewhat irrelevant, Neither relevant nor irrelevant, Somewhat relevant, Relevant, Very relevant)
    \item How useful is the proposed tool expertise for your task? (Very unuseful, Unuseful, Somewhat unuseful, Neither useful nor unuseful, Somewhat useful, Useful, Very useful)
    \item How relevant is the proposed tool expertise for your task? (Very irrelevant, Irrelevant, Somewhat irrelevant, Neither relevant nor irrelevant, Somewhat relevant, Relevant, Very relevant)
    \item How useful is the generated tool for your task? (Very unuseful, Unuseful, Somewhat unuseful, Neither useful nor unuseful, Somewhat useful, Useful, Very useful)
    \item How relevant is the generated tool for your task? (Very irrelevant, Irrelevant, Somewhat irrelevant, Neither relevant nor irrelevant, Somewhat relevant, Relevant, Very relevant)
    \item How much control did you feel you had to modify the tool? (Very insufficient control, Insufficient control, Somewhat insufficient control, Neither sufficient nor insufficient control, Somewhat sufficient control, Sufficient control, Very sufficient control)
    \item How interpretable was the tool generation process? (Very uninterpretable, Uninterpretable, Somewhat uninterpretable, Neither interpretable nor uninterpretable, Somewhat interpretable, Interpretable, Very interpretable)
    \item How would you rate the overall quality of the tool? (Very poor quality, Poor quality, Somewhat poor quality, Neutral, Somewhat good quality, Good quality, Very good quality)
    \item Please briefly describe your comments or reactions to the tool (Free text field)
\end{itemize}

\subsection{Interview}
\label{appendix:interview}
Interview Part 1 (Baseline Comparison Phase):
\begin{itemize}
    \item Output. What was your general impression of the output generated in both conditions?
    \item Goals. What was your impression of the goals generated in [Poppins task]?
    \item Components. What was your impression of the selected tool format and tool expertise generated in [Poppins task]?
    \item Editing process. How did you find the process of trying to improve the system-generated assistance in both conditions?
    \item Overall comparison. What was your overall impression of the systems you used in each condition? Anything that excited you or surprised you?
\end{itemize}

Interview Part 2 (Exploratory Phase):
\begin{itemize}
    \item Output. What was your general impression of the tool that the system generated?
    \item Goals. What was your impression of the goals?
    \item Components. What was your impression of the selected tool format and tool expertise?
    \item Overall comparison. What was your overall impression of this design direction? Anything that excited you or surprised you?
    \item Are there other tasks where you would like to try out our tool? Why?
    \item Would you want to use a tool like this in your everyday life?
    \item Did you have any other thoughts or reflections from the study? Anything that surprised you? Feedback to share?
\end{itemize}

\subsection{Participant Background}
\label{appendix:demographics}
Our participants included 12 women and 5 men; 13 participants were age 18-24, 4 were 25-34. All participants were university students, with 10 undergraduate students and 7 PhD students. They spanned a variety of fields including Computer Science (7), Bioengineering (2), Electrical Engineering (2), Linguistics (1), Chemical Engineering (2), and Chemistry (3). All participants were regular users of large language models, with 14 participants indicating that they use LLMs multiple times in a day and 3 participants used LLMs multiple times in a week. Their usages included Personal use (hobbies, personal projects, entertainment) (14); Professional/work settings (work-related tasks, professional communication, productivity) (15); Educational settings (schoolwork, academic research, coursework) (16); Creative projects (writing, art, design) (9); Technical/development tasks (programming, software development, data analysis) (16); and Social interactions (social media, conversations, community engagement) (3).


\end{document}